\documentclass[sigconf]{acmart}

\usepackage{algorithmic}
\usepackage{algorithm2e}
\usepackage{graphicx}
\usepackage{textcomp}
\usepackage{xcolor}
\usepackage{caption}
\usepackage{subfig}
\usepackage{tablefootnote}
\usepackage{multirow}
\usepackage{xspace}
\usepackage{soul}
\AtBeginDocument{%
  \providecommand\BibTeX{{%
    \normalfont B\kern-0.5em{\scshape i\kern-0.25em b}\kern-0.8em\TeX}}}

\copyrightyear{2023}
\acmYear{2023}
\setcopyright{acmlicensed}\acmConference[ISCA '23]{Proceedings of the 50th Annual International Symposium on Computer Architecture}{June 17--21, 2023}{Orlando, FL, USA}
\acmBooktitle{Proceedings of the 50th Annual International Symposium on Computer Architecture (ISCA '23), June 17--21, 2023, Orlando, FL, USA}
\acmPrice{15.00}
\acmDOI{10.1145/3579371.3589076}
\acmISBN{979-8-4007-0095-8/23/06}

\begin{document}

\title{ImaGen: A General Framework for Generating Memory- and Power-Efficient Image Processing Accelerators}

\author{Nisarg Ujjainkar}
\email{nujjaink@ur.rochester.edu}
\affiliation{%
  \institution{University of Rochester}
  \city{Rochester}
  \state{NY}
  \country{USA}
}

\author{Jingwen Leng}
\affiliation{%
  \institution{Shanghai Jiaotong University}
  \city{Shanghai}
  \country{China}}
\email{leng-jw@sjtu.edu.cn}

\author{Yuhao Zhu}
\affiliation{%
  \institution{University of Rochester}
  \city{Rochester}
  \state{NY}
  \country{USA}
}
\email{yzhu@rochester.edu}


\begin{abstract}

Image processing algorithms are prime targets for hardware acceleration as they are commonly used in resource- and power-limited applications.
Today's image processing accelerator designs make rigid assumptions about the algorithm structures and/or on-chip memory resources. As a result, they either have narrow applicability or result in inefficient designs.


This paper presents a compiler framework that automatically generates memory- and power-efficient image processing accelerators. We allow programmers to describe generic image processing algorithms (in a domain specific language) and specify on-chip memory structures available.
Our framework then formulates a constrained optimization problem that minimizes on-chip memory usage while maintaining theoretical maximum throughput.
The key challenge we address is to analytically express the throughput bottleneck, on-chip memory contention, to enable a lightweight compilation.
FPGA prototyping and ASIC synthesis show that, compared to existing approaches, accelerators generated by our framework reduce the on-chip memory usage and/or power consumption by double digits. ImaGen code is available at: \url{https://github.com/horizon-research/imagen}.

\end{abstract}



\maketitle


\newcommand{\website}[1]{{\tt #1}}
\newcommand{\program}[1]{{\tt #1}}
\newcommand{\benchmark}[1]{{\it #1}}
\newcommand{\fixme}[1]{{\textcolor{red}{\textit{#1}}}}
\newcommand{\fixed}[1]{{\textcolor{orange}{\textit{#1}}}}

\newcommand*\circled[2]{\tikz[baseline=(char.base)]{
            \node[shape=circle,fill=black,inner sep=1pt] (char) {\textcolor{#1}{{\footnotesize #2}}};}}

\ifx\figurename\undefined \def\figurename{Figure}\fi
\renewcommand{\figurename}{Fig.}
\renewcommand{\paragraph}[1]{\textbf{#1} }
\newcommand{\figline}{{\vspace*{.05in}\hline}}

\newcommand{\Sect}[1]{Sec.~\ref{#1}}
\newcommand{\Fig}[1]{Fig.~\ref{#1}}
\newcommand{\Tbl}[1]{Tbl.~\ref{#1}}
\newcommand{\Equ}[1]{Equ.~\ref{#1}}
\newcommand{\Apx}[1]{Apdx.~\ref{#1}}
\newcommand{\Alg}[1]{Algo.~\ref{#1}}

\newcommand{\specialcell}[2][c]{\begin{tabular}[#1]{@{}c@{}}#2\end{tabular}}
\newcommand{\note}[1]{\textcolor{red}{#1}}

\newcommand{\proj}{\textsc{Crescent}\xspace}
\newcommand{\mode}[1]{\underline{\textsc{#1}}\xspace}
\newcommand{\sys}[1]{\underline{\textsc{#1}}}

\newcommand{\no}[1]{#1}
\renewcommand{\no}[1]{}
\renewcommand{\hl}[1]{#1}
\newcommand{\RNum}[1]{\uppercase\expandafter{\romannumeral #1\relax}}

\def\cA{{\mathcal{A}}}
\def\cF{{\mathcal{F}}}
\def\cN{{\mathcal{N}}}
\def\bh{{\mathbf{h}}}
\def\bp{{\mathbf{p}}}


\graphicspath{{figs/}}

\section{Introduction}

Image processing has become ever more important with a plethora of emerging visual computing domains such as Augmented/Virtual Reality, computational photography, and smart cameras. These application domains all present stringent resource and power constraints, leading to many research efforts in building specialized accelerators for image processing~\cite{gan2021eudoxus,redgrave2018pixel,rigel,ivs,paisp,ranganathan2021warehouse}.
Manually building accelerators, however, is not only time-consuming, error-prone, but also relies heavily on empirical heuristics that do not always deliver optimal designs.




A recent trend is automatically generating accelerators from high-level algorithm descriptions~\cite{hegarty2014darkroom, whatmough2019fixynn, chi2018soda}.
Prior approaches to generating image processing accelerators either have narrow applicability or yield inefficient designs --- for two main reasons (\Sect{sec:motivation}).
First, they optimize for simple, single-consumer algorithms where each producer stage has only one consumer.
When facing multiple-consumer algorithms such as unsharp filtering~\cite{hegarty2014darkroom} and denoising~\cite{chi2018soda}, they either have to artificially transform the multiple-consumer algorithm to a single-consumer arrangement, which increases the on-chip memory usage, or increase the total on-chip memory accesses, which increases the power consumption.

Second, there is a large, algorithm-dependent trade-off space between on-chip memory requirement and power consumption that prior work fails to explore.
This is because prior work assumes one single memory structure and, critically, use the same memory structure for \textit{all} algorithms and for \textit{all} stages in an algorithm.
For instance, FixyNN~\cite{whatmough2019fixynn} could generate designs using only single-port SRAMs, and SODA~\cite{chi2018soda} could generate designs using only FIFOs (dual-port SRAMs).
The actual design space is much larger: given an algorithm with $N$ stages and $M$ memory structures, there are $M^N$ design points, each providing a unique power-vs-area trade-off.



This paper proposes a compiler framework that generates memory- and power-efficient accelerators (in the form of synthesizable RTL) for image processing (\Sect{sec:compiler}).
Instead of artificially restricting algorithm and/or on-chip memory structures, we allow specifying generic algorithms and memory configurations (in terms of size and number of ports).
Given the algorithm and hardware specifications, our compiler formulates a constrained optimization problem that, while maintaining theoretically maximum throughput, minimizes the on-chip memory usage and reduces total power consumption.


A key challenge we address is to generate accelerators that consistently deliver theoretically maximum throughput (frame rate) \textit{for every frame}; after all, saving on-chip area and power consumption is of little use when an image processing accelerator has a low frame rate.
The central difficulty is to analytically express the throughput bottleneck, i.e., on-chip memory contention, which involves set counting and is incompatible with numerical optimizations. We leverage the data access pattern of stencil operations to transform set counting into equivalent, arithmetic operations that are amenable to numerical optimizations (\Sect{sec:opt}).

Building on top of the optimization formulation, we propose to judiciously coalesce multiple lines in a line buffer into a single memory block to further reduce on-chip memory consumption. We show that this technique amounts to a static rewriting of the algorithm Directed Acyclic Graph (DAG) and is naturally integrated into our compiler framework (\Sect{sec:lc}).


We show that our optimization problem is an Integer Linear Programming (ILP), which has efficient solvers. As a result, our compiler is lightweight; it generates synthesizable RTL for common image processing algorithms in milliseconds.
We use our framework to generate a wide variety of image processing accelerators, which we evaluate using both an ASIC flow and a Xilinx Spartan-7 FPGA board. Across different input sizes, accelerators generated by our framework reduce on-chip memory usage and power by up to 86.0\% and 62.9\%, respectively, when compared to designs generated by prior methods, including Darkroom~\cite{hegarty2014darkroom}, SODA~\cite{chi2018soda}, FixyNN~\cite{whatmough2019fixynn}.

We use our framework to perform
a Design Space Exploration (DSE) that explores diverse memory configurations to generate Pareto-optimal designs.
We show that the area-vs-power trade-off varies with algorithms, an algorithm-specific design space exploration that our framework uniquely enables.


In summary, this paper makes the following contributions:

\begin{itemize}
  \item We propose a compiler framework that generates memory- and power-efficient image processing accelerators given generic algorithm and on-chip memory specifications. The accelerators guarantee theoretically maximum throughput through constrained optimization.
  \item We propose a line-coalescing algorithm that coalesces multiple line-buffer lines into one memory block to further reduce on-chip memory usages.
  \item Accelerators generated by our compiler consume less on-chip memory and power compared to those generated using existing tools. Our compiler is integrated into a DSE process, which reveals algorithmic-dependent area-vs-power trade-offs that prior tools are unable to explore. 

\end{itemize}

%
%
%

\section{Background}
\label{sec:background}



Image processing pipelines consist of computation stages that operate on regular 2D pixel arrays.
Each stage performs a stencil operation, which operates on a window of input pixels to generate an output pixel.
The stencil window moves in a raster scan order.
An end-to-end algorithm usually cascades multiple stages.
Each stage generates an intermediate 2D image read by (potentially multiple) consumer(s). Common image processing algorithms include in-camera image signal processing~\cite{hegarty2014darkroom} and High Dynamic Range imaging using burst photography~\cite{hasinoff2016burst}.

\paragraph{Scope.}
Our goal is \textit{not} a generic stencil accelerator that runs multiple algorithms. Rather, we focus on accelerators that are specialized for a given algorithm. This is common in both 1) FPGA-based acceleration systems, where the FPGA can be re-programmed for a given algorithm, and 2) low-power ASICs as demonstrated in Image Signal Processors in modern cameras (e.g., Arm Mali~\mbox{\cite{malic55}} and Qualcomm Spectra~\mbox{\cite{qcspectra}}), embedded computer vision processors~\mbox{\cite{feng2019asv, zhu2018euphrates, mahmoud2017ideal,  whatmough2019fixynn}}, and robotics accelerators~\mbox{\cite{suleiman2019navion, sacks2018robox, murray2016microarchitecture}}, where extreme efficiency requires algorithm-specific accelerators.




\begin{figure}[t]
    \centering
    \subfloat[Cycle t.]{
        \label{fig:lb1}
        \includegraphics[scale=0.4]{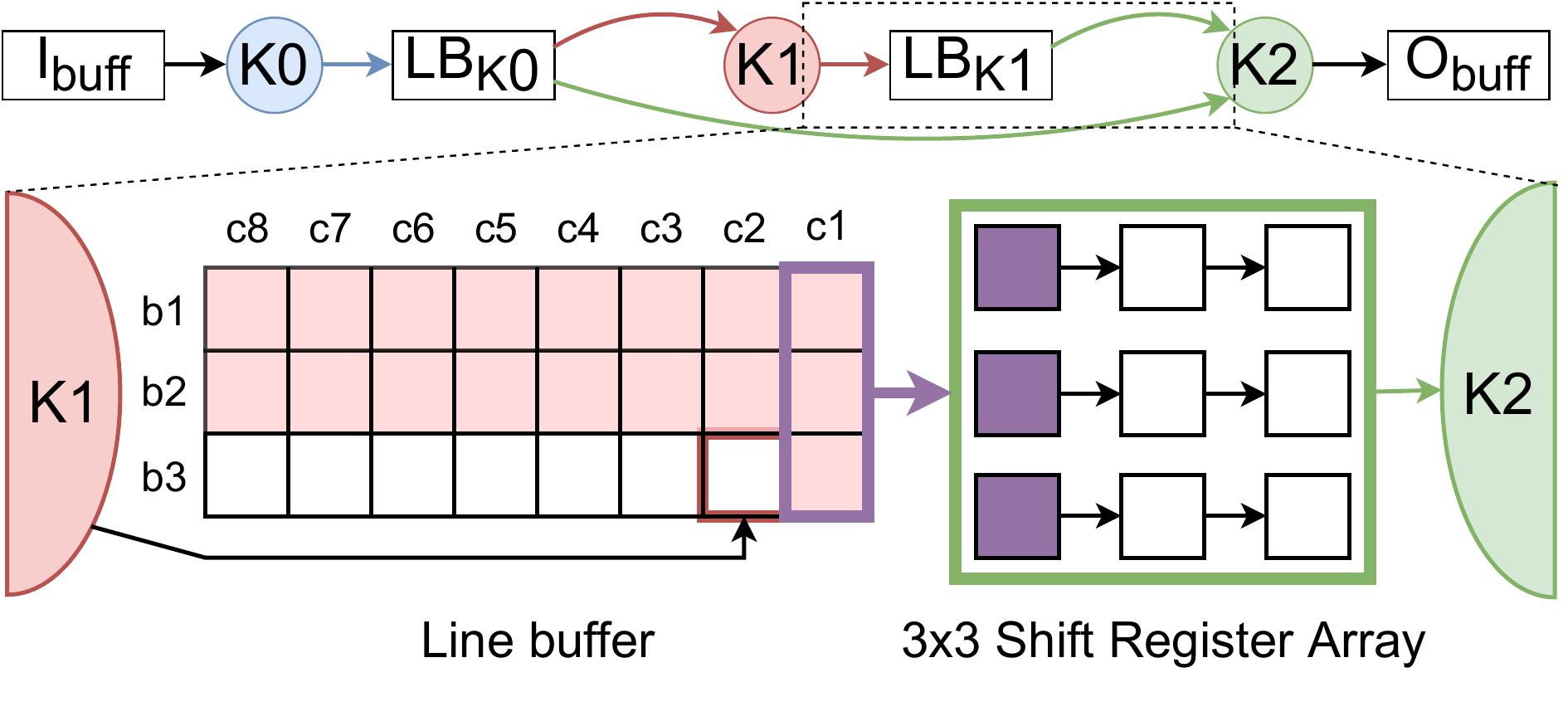}
    }
    
    \subfloat[Cycle t+1.]{
        \label{fig:lb2}
        \includegraphics[scale=0.4]{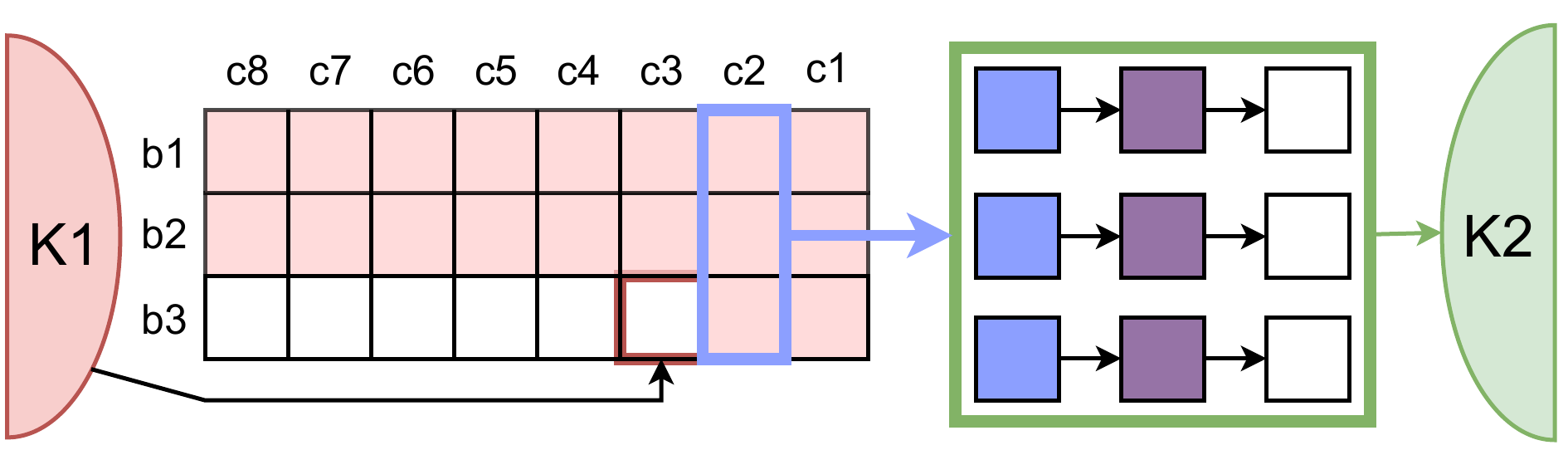}
    }
    
    \subfloat[Cycle t+2.]{
        \label{fig:lb3}
        \includegraphics[scale=0.4]{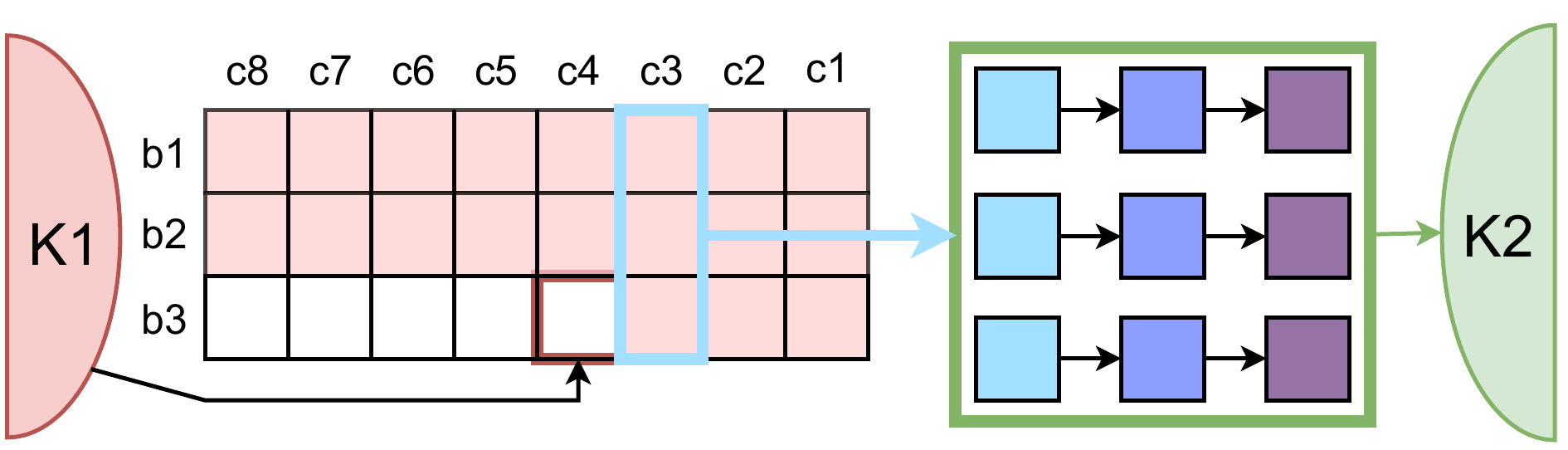}
    }
    
    \subfloat[Cycle t+7.]{
        \label{fig:lb4}
        \includegraphics[scale=0.4]{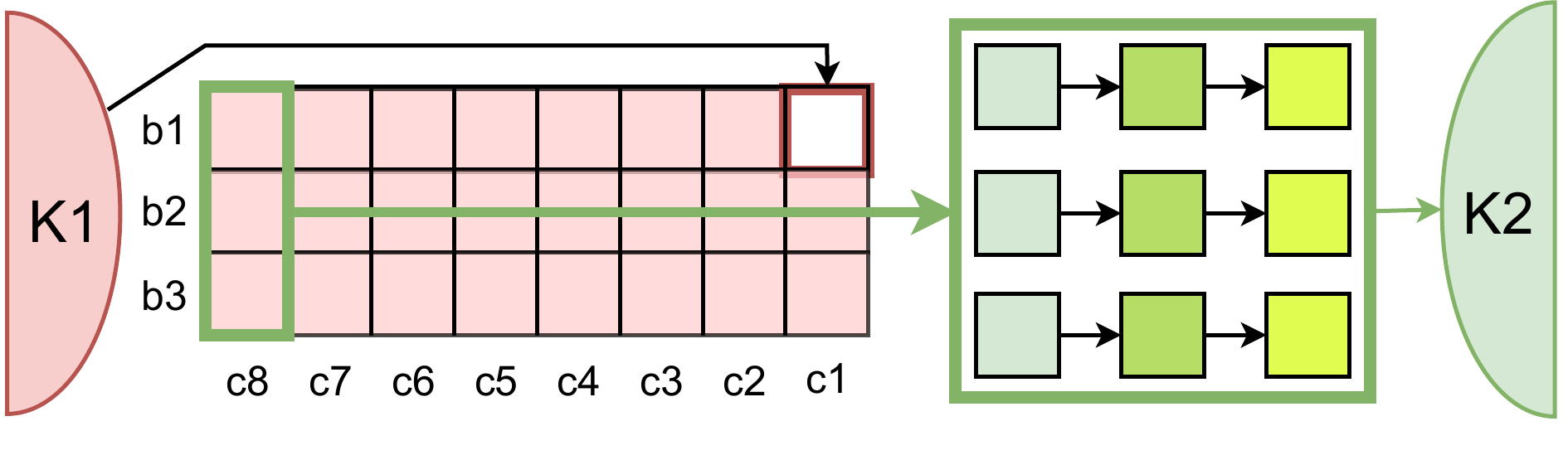}
    }
    \caption{A line-buffered accelerator. Every stage has a dedicated line buffer (and the associated shift register array) to write to. We use the line buffer after stage $K1$ to illustrate the line-buffering operations.
    The line buffer stores three-pixel rows, each stored in a two-port SRAM. As the producer writes the second element in \textsf{b3}, pixels in column \textsf{c1} are moved to the shift register array. After three cycles the shift register array contains the data from a $3 \times 3$ stencil window and $K2$ starts. When $K1$ finishes writing to \textsf{b3}, it writes to \textsf{b1}, since pixels there will no longer be needed by the consumer.
    }
    \label{fig:line_buffer}
\end{figure}


\paragraph{Image Processing Accelerators.}
\Fig{fig:line_buffer} shows a typical image processing accelerator using a simple pipeline as an example. Each algorithm stage is mapped to a dedicated hardware stage.
The input and output stages ($K0$ and $K2$ here) interface with input/output buffers that communicate with off-chip memory. As extensively studied before, those buffers are usually doubled buffered with high access bandwidth, and are \textit{not} the focus of this paper.

Stages communicate intermediate data through another set of on-chip buffers, which make up the majority of the on-chip memory usage.
In particular, each producer stage has a dedicated buffer to write its intermediate data to; all its consumers read from that buffer. This is consistent with all image processing accelerator designs~\cite{hegarty2014darkroom, chi2018soda, whatmough2019fixynn, bagni2017demystifying}.

Ideally, all the inter-stage data communication traffic should be fulfilled entire on-chip; otherwise, off-chip memory accesses would stall the pipeline, which requires complicated hardware logic for dynamic scheduling, introduces non-deterministic frame rates, compromises peak throughput, and consumes high power.
A naive approach is to buffer all the intermediate data between stages on-chip.
This comes with two downsides. First, the intermediate data could be large in size and exceeds a typical on-chip memory capacity. For instance, each $1080p$ image passed around between stages consume 6 MB of data.
Second, it artificially forces the consumer to wait until the producer finishes generating an entire image.


\paragraph{Line Buffer.} A common strategy to address these issues is to use a special on-chip buffer structure called ``line buffer''. The key observation is that each pipeline stage, at any time, operates only on data in a small, local window. Therefore, a consumer stage can start as soon as the data in a stencil window is available, essentially consuming pixels incrementally as they are generated by the producer. A pixel in the buffer can be over-written when it is no longer needed by the consumer (i.e., the stencil window has gone completely past the pixel), reducing the on-chip memory requirement.

Consider the producer-consumer pair $K1$ and $K2$ in \Fig{fig:line_buffer}, where $K2$ operates on a $3\times 3$ stencil window from the output of $K1$ in every cycle. To support this data communication pattern, the hardware uses a line buffer that stores three rows of pixels generated by $K1$; the line buffer is connected to a $3 \times 3$ shift register array, which holds the data in a stencil window and is read by $K2$.

The producer starts writing from the first row, one pixel at a cycle.
As soon as the producer finishes producing two rows plus one element (\Fig{fig:lb1}), the consumer can move the first column of pixels (\textsf{c1} here) from the line buffer to the shift register array.
In the next cycle (\Fig{fig:lb2}), as the producer writes to the third element of \textsf{b3}, pixels in column \textsf{c2} are moved to the shift register array. After three cycles (\Fig{fig:lb3}), the shift register array contains the data from a stencil window, at which point $K2$ can start to produce its output.
Once $K1$ finishes writing to \textsf{b3}, it will write to, instead of a new row, the first row in the line buffer (\textsf{b1}), overwriting data in \textsf{b1}, because pixels there are no longer needed by the consumer (\Fig{fig:lb4}). As a result, the line buffer has to store only three rows of pixels.

\paragraph{Implementation.} In the actual hardware implementation of a line buffer, one would use 3 separate SRAMs, each storing one row of pixels.
At any give cycle, all three SRAMs are being read from and there is one SRAM that is also being written to. The SRAM being written to will rotate. As a result, all the SRAMs must have at least two ports in this example.
Note that when an image row is larger than SRAM block size, the row can be split into multiple SRAM blocks without changing the operating principle of line buffers.


\section{Motivation}
\label{sec:motivation}

\begin{figure}[t]
    \centering
    \includegraphics[width=\linewidth]{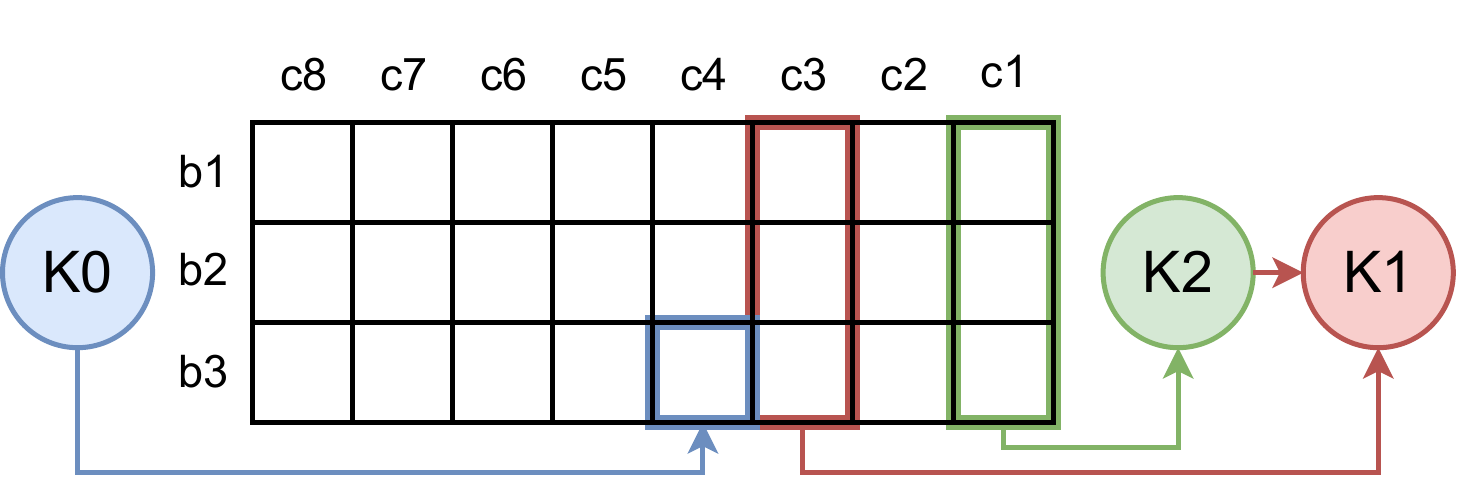}
    \caption{Illustration of a line buffer with multiple consumers. $K0$ is the producer who is writing to cell \textsf{c4} in block \textsf{b3}. $K1$ and $K2$ are consumers that are reading columns \textsf{c3} and \textsf{c1} respectively. Each line is in an SRAM block. Assuming the common setting where each SRAM has two ports, the pipeline stalls since \textsf{b3} has to serve three accesses.}
    \label{fig:mc}
\end{figure}

Accelerator design decisions must be made according to the specific algorithm pipeline and the memory resources available.
Prior work makes rigid assumptions about the algorithm structures and/or memory resources available. As a result, they either have narrow applicability or result in inefficient designs. We summarize prior work in \Tbl{tab:prior} and elaborate next.

\subsection{Algorithmic Limitations}

The basic design in \Sect{sec:background} assumes dual-port SRAMs and ``single-consumer'' pipelines, where each producer has only one consumer.
Multiple-consumer pipelines such as unsharp mask~\cite{hegarty2014darkroom} and denoise2D~\cite{chi2018soda}, where multiples consumers read the output from a producer, challenge the simple design. With multiple consumers, one line is accessed by multiple hardware stages. As a result, there could be more accesses to an SRAM block than there are ports, leading to pipeline stalls.


\begin{table}[t]
\centering
\caption{Prior work makes rigid assumptions about the memory structures, which leads to sub-optimal designs, and/or applies to only specific forms of algorithms.}
\label{tab:prior}
\resizebox{\columnwidth}{!}{
\renewcommand*{\arraystretch}{1.1}
\renewcommand*{\tabcolsep}{2pt}
\begin{tabular}{ccccc}
\toprule[0.15em]
 ~ & \textbf{Darkroom~\cite{hegarty2014darkroom}} & \textbf{FixyNN~\cite{whatmough2019fixynn}} & \textbf{SODA~\cite{chi2018soda}} & \textbf{Ours} \\ 
\midrule[0.10em]
\specialcell{On-chip memory\\ assumption} & Dual port & Single port & \specialcell{Dual port\\ (FIFO)} & \textbf{Generic} \\ \hline
\specialcell{Algorithm\\ applicability} & Single consumer & Single consumer & Generic & \textbf{Generic} \\
\bottomrule[0.15em]
\end{tabular}
}
\end{table}

Consider the algorithm in \Fig{fig:line_buffer}, where $K0$ has two consumers $K1$ and $K2$. The line buffer after $K0$ is simultaneously accessed by three stages (the producer $K0$ and the two consumers). \Fig{fig:mc} shows a naive line buffer design, where $K0$ is writing to (\textsf{b3}, \textsf{c4}), and $K1$ and $K2$ are reading columns \textsf{c3} and \textsf{c1}, respectively. As a result, three different stages are accessing the block \textsf{b3}. If there are only two ports in each SRAM block, \textsf{b3} will not be able to handle all three accesses.
Simply increasing the number of SRAM ports is area-inefficient as SRAM area increases quadratically with the number of ports~\cite{weste2015cmos}.
Prior work attempts to support multi-consumer pipelines in primarily two ways, each coming with its own downsides, which we explain next.

\begin{figure}[t]
    \centering
        \includegraphics[width=\columnwidth]{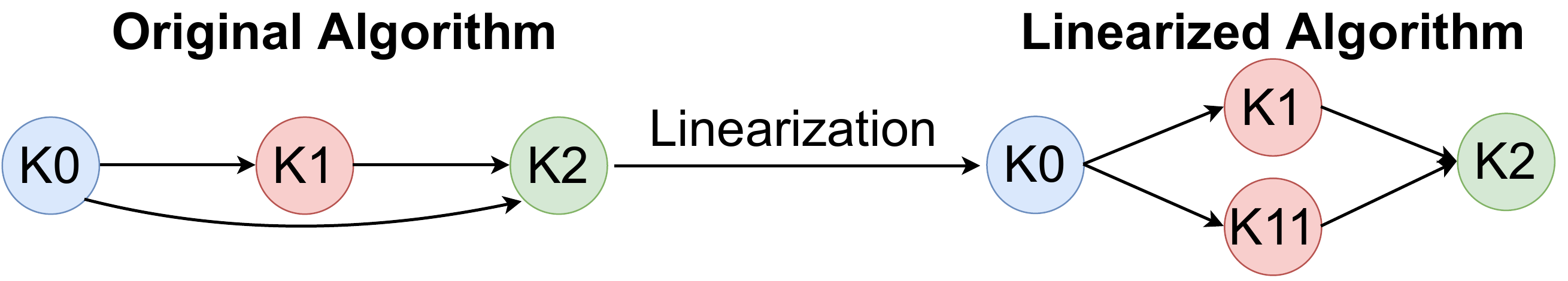}
    
    \caption{
    An example of algorithm linearization. We need an additional line buffer after $K11$. Note that even though $K1$ and $K11$ are both consumers of $K0$, they consume data in exactly the same pattern and act effectively as one consumer.}
    \vspace{-5pt}
    \label{fig:linear_a}
\end{figure}
  
\paragraph{Algorithm linearization.} One can transform an algorithm that has multiple-consumer stages into another functionally identical algorithm with only single-consumer stages, a process dubbed ``linearization'' by Darkroom~\cite{hegarty2014darkroom}.

We use \Fig{fig:linear_a} to illustrate linearization.
To linearize the pipeline, we add a dummy stage $K11$ between $K0$ and $K2$. Each cycle, $K11$ reads data from $K0$ in \textit{exactly the same pattern} as $K1$ but performs no computation on the data.
Essentially, the sole purpose of $K11$ is to simply relay data from $K0$ to $K2$.
As a result, $K2$ reads data from $K11$ instead of directly reading from $K0$. Critically, even though $K0$ still has two consumes $K1$ and $K11$, the two consumers read data from $K0$ in exactly the same pattern every cycle, so they effectively act as a single consumer without requiring additional memory port.


The downside of this approach is that the linearized algorithm increases on-chip memory usage than the original algorithm, because each dummy stage requires a dedicated line buffer. 
In this case, the additional line buffer associated with $K11$ buffers the same data that $K1$ does and is redundant.
 
\begin{figure}[t]
    \centering
    \subfloat[An FIFO implementation to support a single consumer ($K1$)]{
        \label{fig:fifo1}
        \includegraphics[width=\columnwidth]{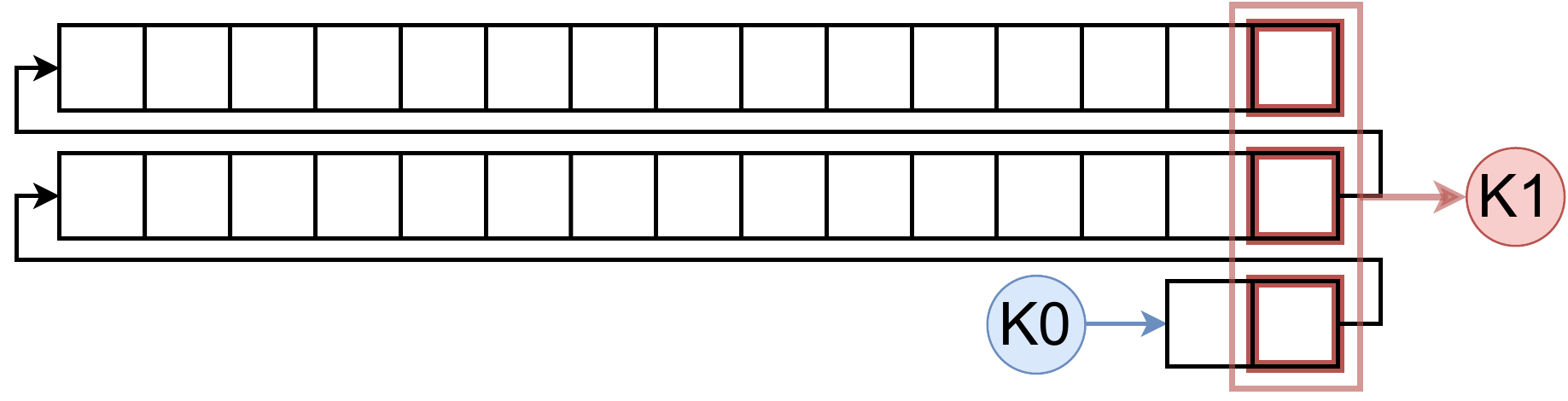}
    }
    
    \subfloat[An FIFO implementation to support multiple consumers ($K1$ and $K2$).]{
        \label{fig:fifo2}
        \includegraphics[width=\columnwidth]{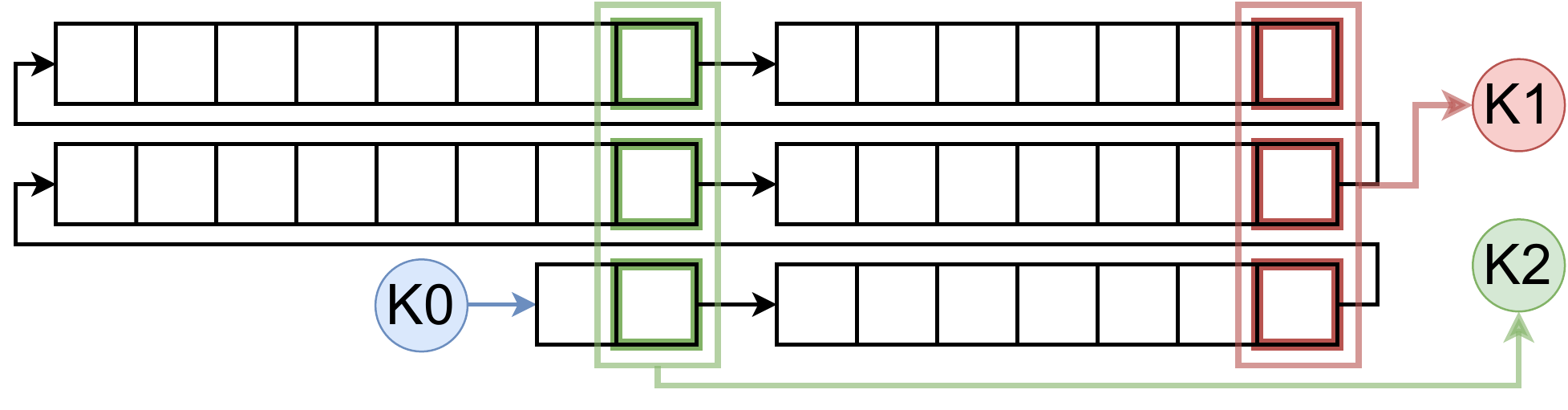}
    }
    \caption{Line buffer implementation using FIFO. Figure \protect\subref{fig:fifo1} shows the single consumer case and \protect\subref{fig:fifo2} shows a multiple consumer case, assuming each memory block has two ports. In the multiple consumer case, to accommodate both consumers each FIFO is split into two smaller FIFOs. A FIFO is usually implemented as a dual-port SRAM block. The line being written to by the producer usually holds only a few elements (2 here) and, thus, can be implemented as a shift register using DFFs to save on-chip memory usage.}
    \vspace{-5pt}
    \label{fig:fifo}
\end{figure}
  
\paragraph{Splitting line buffer.} Alternatively, one can split a line buffer that has multiple consumers into several smaller line buffers, each serving only one consumer.
An implication of this approach is that data from one line buffer must be transferred to another, which requires each line buffer to be realized as a FIFO. This approach is exemplified by SODA~\cite{chi2018soda} and line-buffered designs from Xilinx ~\cite{bagni2017demystifying}.

\Fig{fig:fifo1} shows FIFO-based line buffer implementation when there is only a single consumer. Usually a FIFO is implemented using a dual-port SRAM/BRAM block. Therefore, every cycle there is one read and one write access to every memory block. When two consumers are trying to access the line buffer, each FIFO must be split into two FIFOs, each in a separate memory block, to accommodate both accesses; this is illustrated in \Fig{fig:fifo2}.
Note that if a FIFO becomes very small (e.g., a few elements), which is typically the case for the line that the producer is writing to, it can be implemented as a shift register using DFFs to reduce memory usage.

The downside of FIFOs is high energy consumption, because the nature of FIFO dictates that every cycle there would \textit{always} be two accesses to each SRAM block, whereas in the classic implementation only one of three lines have to serve two accesses; other lines have only one access each cycle. In our FPGA measurement (see \Sect{sec:exp} for details), BRAMs with two accesses per cycle consume about 35\% more power than BRAMs with only one access per cycle.

\subsection{Hardware Limitations}

A fundamental limitation of prior approaches is that they use the same form of memory for \textit{all} the algorithms and for \textit{all} the line buffers in an algorithm, e.g., dual-port SRAM in Darkroom~\cite{hegarty2014darkroom} or single-port SRAM in FixyNN~\cite{whatmough2019fixynn}.
The actual design space, however, is much larger.
For an algorithm with $N$ stages where a line buffer has $M$ implementation options, there are $M^N$ design points.
Navigating a large design space governed by memory resources is especially important in ASIC designs, where, unlike FPGAs where memory resources are fixed, one has the flexibility to customize memories (e.g., size and ports) given area and/or power targets.


Critically, there exists a power-vs-area trade-off when navigating such a large design space.
For instance, increasing the number of memory ports increases the per-SRAM area and power consumption but also reduces the number of SRAM blocks needed. 
As we will show in \Sect{sec:res:dse}, the exact Pareto-optimal frontier varies across algorithms, a design space exploration that is not possible with existing approaches.

Finally, it is worth noting that prior approaches could not generate any hardware design at all when the available memory resource does not meet their requirement.
For instance, the FIFO approach by SODA~\cite{chi2018soda} assumes dual-port memories.
It thus does not work when only single-port memories are available, further highlighting the need to consider arbitrary memory configurations, which our framework offers.


\section{Framework Overview}
\label{sec:compiler}

\begin{figure}[t]
    \centering
    \includegraphics[width=.8\columnwidth]{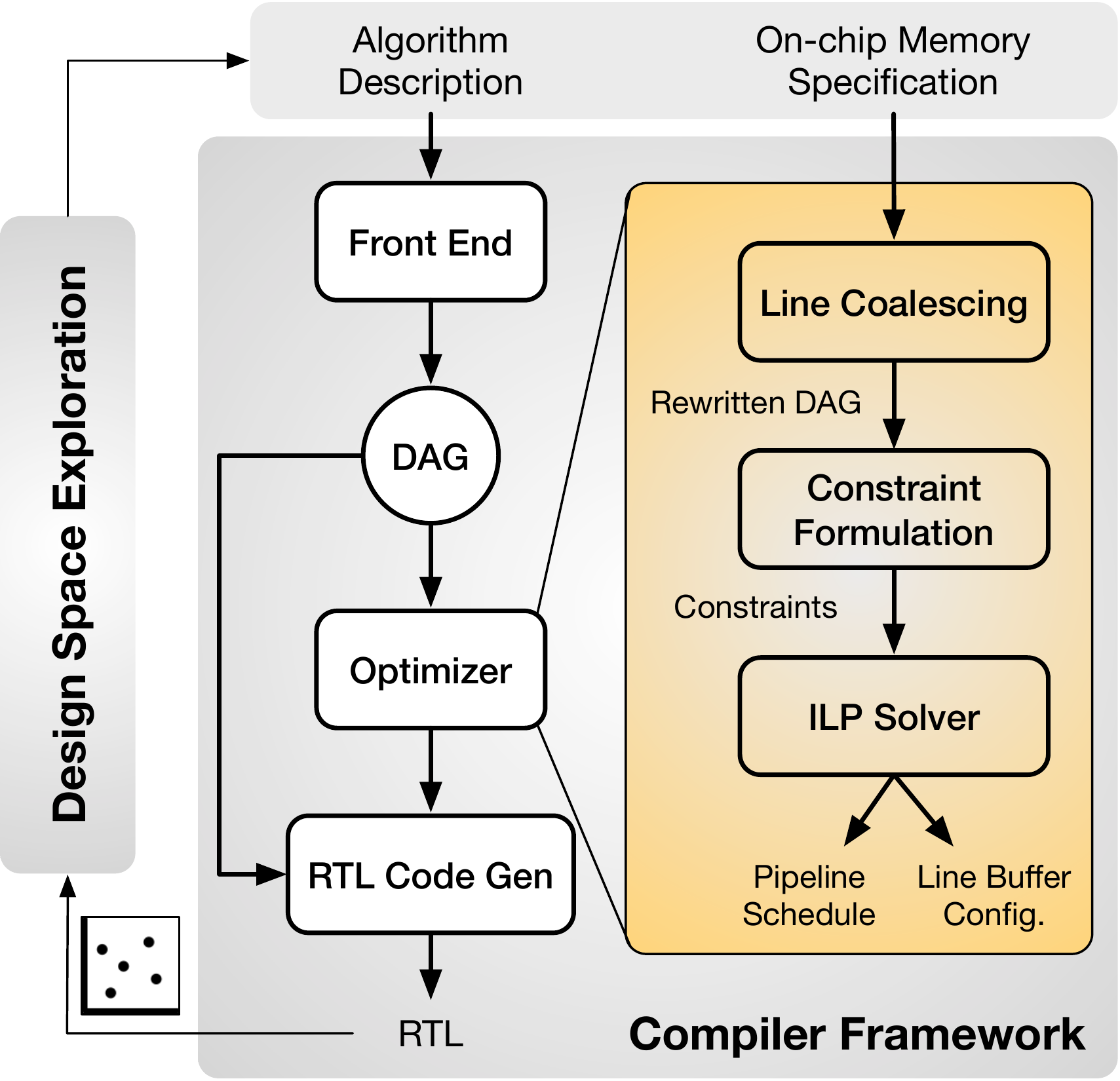}
    \caption{Compiler framework, which generates synthesizable Verilog code given an image processing algorithm and on-chip memory specifications. The framework can be integrated into a DSE tool to generate Pareto-optimal designs.}
    \label{fig:compiler}
    \vspace{-5pt}
\end{figure}

\Fig{fig:compiler} shows the overall workflow of our framework, which takes an image processing pipeline described in a Domain Specific Language (DSL) and the description of available memory resources (i.e., sizes and number of ports) and generates synthesizable Verilog code.


\paragraph{Front End.} Literature is rich with DSLs that express image processing pipelines~\cite{hegarty2014darkroom, ragan2013halide}, which is \textit{not} the focus of this paper. For simplicity, we use a DSL similar to Darkroom~\cite{hegarty2014darkroom}.
The code below shows the algorithm in \Fig{fig:line_buffer}. Each stage is defined inside the \texttt{im} block.
\texttt{input} and \texttt{output} denote input and output stages of the pipeline, for which off-chip memory accesses are permitted.
The front-end parses an algorithm to a DAG as the intermediate representation. Each DAG node is a pipeline stage, and each edge connects a producer-consumer pair. The stencil window sizes are encoded in DAG nodes.

\begin{figure}[h]
    \centering
    \includegraphics[width=\linewidth]{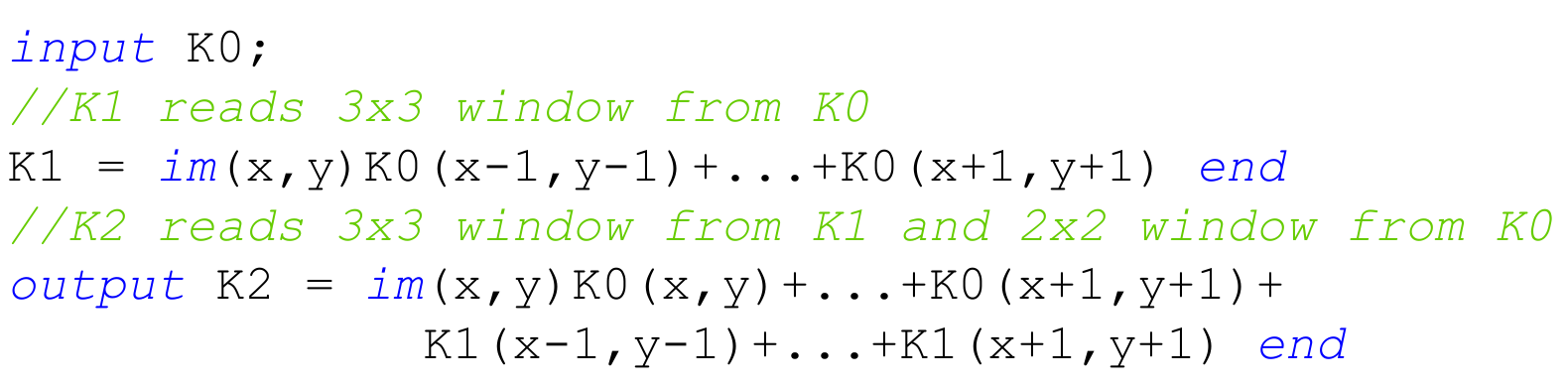}
    \label{fig:code}
\end{figure}

\paragraph{Optimizer.} The core contribution of our work is the optimizer, which takes a DAG and the memory resource specifications to generate accelerator pipeline schedule and on-chip memory configurations, which are then used in generating synthesizable RTL code for by the code generator.

The optimizer forms a constrained optimization problem that minimizes the on-chip memory usage (\Sect{sec:opt}). Unlike conventional line buffer synthesizers~\cite{chi2018soda, whatmough2019fixynn, hegarty2014darkroom}, the compiler exploits opportunities to coalesce image rows into the same buffer through a simple DAG re-writing, further reducing on-chip memory usage (\Sect{sec:lc}).


\paragraph{RTL Code Generation.} 
Given the pipeline schedule and the line buffer configuration, the code generator generates synthesizable RTL.
The generated code has compute units that execute the stencil operations, memory blocks that implement line buffers, and control logic to sequence the hardware.

It is worth noting that the code geeneration is largely a \textit{mechanical translation} of arithmetic operations in each pipeline stage to RTL code and, thus, is \textit{not} a contribution of this paper.
In fact, any existing HLS tool (e.g., Vivado HLS) can be used for code generation: one can use our optimizer to generate the optimal line buffer configuration, which is then used in the HLS code to annotate the memory sizes.
To our best knowledge, today’s HLS tools such as Vivado HLS require programmers to explicitly specify line buffer sizes, which our optimizer automatically generates.


 
\no{\paragraph{Advantages.} Compared to prior work, our framework provides two advantages. First, it does not constrain algorithm and hardware; instead, it generates memory-optimal accelerator \textit{given} the memory and algorithm specification. In contrast, prior work works only for specific memories or algorithms. Second, because our framework accepts generic memory specifications, it can be easily integrated into a design space exploration tool that sweeps the memory configurations to generate Pareto-optimal designs (\Sect{sec:res:dse}).}

\section{Generating Line-Buffered Pipelines}
\label{sec:opt}

\begin{table}[t]
\centering
\caption{List of symbols used in our formulation. Subscripts $i, j, p, c$ are used in denoting pipeline stages; $p$ and $c$, in particular, denote a producer and consumer, respectively. Blackboard-bold symbols $\mathbb{N}, \mathbb{C}, \mathbb{A}$ denote sets.}
\label{tab:symb}
\resizebox{\columnwidth}{!}{
\renewcommand*{\arraystretch}{1.1}
\renewcommand*{\tabcolsep}{4pt}
\begin{tabular}{cl}
\toprule[0.15em]
  \textbf{Symbols} & \textbf{Meaning}\\ 
\midrule[0.10em]
  $N$ & Total number of stages in the pipeline \\
  $LB_i$ & Size of the line buffer associated with stage $i$ \\
  $S_i$ & Start cycle of a stage $i$ \\
  $\mathbb{N}_i$ & Set of stages accessing the line buffer of stage $i$ \\
  $SH_i$ & Stencil height of stage $i$\tablefootnote{A consumer stage can access data from multiple producers and, thus, can have multiple stencil heights. We omit the producer stage from the symbol for simplicity, but the producer stage is evident in each context where $h_i$ appears.} \\
  $B_{l,t}$ & Total number of accesses to a line $l$ at cycle $t$ \\
  $P$ & Number of SRAM ports \\
  $W$ & Width of the input image to each stage\tablefootnote{\hl{Our current system, as is, can deal with stages without padding, in which case the input image size is different across stages and is trivially calculated given the stencil window size. For the simplicity of the exposition we assume padding and use the same $W$ in the paper.}}\\
  $\mathbb{C}_p$ & Set of consumer stages of a producer stage $p$ \\
  $\mathbb{A}_{i,t}$ & Set of lines stage $i$ is accessing at cycle $t$\tablefootnote{Similar to $SH_i$, $\mathbb{A}_{i, t}$ here is tied to a particular producer of $i$, which could have multiple producers. The producer is omitted in the symbol for simplicity, but should be evident given the context.} \\ 
  $L_{i,t}$ & The first line that a stage $i$ is accessing at cycle $t$ \\
\bottomrule[0.15em]
\end{tabular}
}
\end{table}

We first describe the intuition behind our general idea (\Sect{sec:opt:intuition}), followed by a rigorous optimization formulation (\Sect{sec:opt:prob}). We then discuss how on-chip memory contention, the key to our optimization formulation, is modeled (\Sect{sec:opt:hw}), followed by a technique to eliminate redundant hardware constraints (\Sect{sec:opt:cp}). In the end, we show that our formulation amounts of an ILP problem (\Sect{sec:opt:solver}).

\subsection{General Idea and Intuition}
\label{sec:opt:intuition}

\paragraph{Objective.} Our goal is to minimize the total on-chip memory size while \textit{maintaining the theoretically maximum throughput}, which is quantified by the number of pixels generated per cycle. The theoretically maximum throughput is fundamentally limited by the amount of functional units the hardware can afford to have. Like virtually all prior work~\cite{chi2018soda, whatmough2019fixynn, hegarty2014darkroom}, we assume that the theoretically maximum throughput is one pixel per cycle. Improving the throughput simply amounts to increasing the compute resources, which is \textit{not} the focus of this paper. Note, however, that the one-pixel-per-cycle assumption is reasonable for real-world applications. Assuming a 100 MHz clock frequency, producing one pixel per cycle is equivalent to providing a 50 frames per second (FPS) frame rate for $1080p$ images, sufficient for real-time operations.

We emphasize that the accelerator must \textit{consistently} deliver the prescribed throughput across frames. It is unacceptable if some frames are lower that others even if the average frame rate is desirable, because a varying frame rate presents a sluggish user experience.
To ensure a consistent throughput, the accelerator pipeline must not stall (once a pipeline starts it never stalls until all the input pixels finish), which translates to meeting three requirements:

\vspace{5pt}

\noindent \textit{\boxed{R1}~\underline{Data dependency (causality)}: any pixel, before can be read by a consumer, must already be generated by the producer and is available in the line buffer};
\vspace{5pt}

\noindent \textit{\boxed{R2}~\underline{\hl{No intermediate off-chip memory access}}: a pixel is evicted from a line buffer only when it is no longer needed by any of its consumers (to avoid DRAM accesses later)};
\vspace{5pt}

\noindent \textit{\boxed{R3}~\underline{No on-chip memory access stall}: at any cycle, the number of accesses to any on-chip memory block must be no more than the number of ports.}

\vspace{5pt}

\paragraph{Solution Intuition.} Intuitively, a generic solution to meeting both \hl{memory requirements} is to delay the start of certain consumers.
For instance, when two consumers of the same producer are allowed to start at the same cycle, each memory block has to serve two simultaneous read accesses. Now if one consumer is delayed by $W$ cycles where $W$ is the width of an image, the two consumers will be reading from two different memory blocks (image rows), avoiding memory-port contention. Delaying the start of a consumer also providing more time for the producer to generate intermediate data.

Delaying consumers, however, increases the line buffer size, because an element is evicted from the line buffer only when it is no longer needed by \textit{any} consumer; delaying a consumer would mean that data will have to stay in the line buffer for longer, increasing the line buffer size.
Thus, the central challenge is how to optimally shift: how to schedule different pipeline stages to minimize total line buffer size while meeting the data and hardware constraints.

%


\subsection{Optimization Formulation}
\label{sec:opt:prob}


Formally, the job of our hardware generator can be described in a constrained optimization formulation:
\begin{subequations}
\begin{align}
  \min_{\phi}~&LB(\phi)=\sum_{i=0}^{N-1}LB_i(\phi), \nonumber\\
  & where~\phi = \{S_i\}, i \in [0, 1, \cdots, N-1] \label{eq:objf}\\
  s.t.~~~& \forall (p, c)~~S_c - S_p \geq (SH_c - 1)\times W + 1 \label{eq:cc},\\
       & \forall l \forall t~~ B_{l,t}(\phi)\leq P. \label{eq:chw}
\end{align}
\label{eq:opt}
\end{subequations}
\paragraph{Optimization Objective.}  \Equ{eq:objf} states the optimization objective. $\phi$, the schedule, denotes the collection of the start cycles of all the stages $\{S_i\}$ ($i $ is an integer between $0$ and $N-1$, where $N$ is the number of pipeline stages). $LB(\phi)$ denotes the total line buffer size, which is the sum of the $N$ individual line buffer sizes. Recall from \Fig{fig:line_buffer} that each stage is associated with a line buffer, so the number of line buffers is the same as the number of pipeline stages, $N$.

The size of each line buffer is dictated by the start cycles of the producer stage and the consumer stages. For instance, given a producer-consumer pair $(p, c)$ in the pipeline and their start cycles $S_p$ and $S_c$, there is a $(S_c - S_p)$-cycle delay between the consumer and the producer.
According to \boxed{R2} above, the line buffer must have the ability to buffer at least $(S_c - S_p)$ pixels before the consumer starts, because each pixel can be removed from the line buffer only after it has been consumed.



Considering that there could be multiple consumers of the same producer and any element can be removed from the line buffer only after the \textit{last} consumer finishes reading it, the line buffer size of a particular stage $p$ is:

\begin{equation}
	LB_p = \left\lceil \frac{\max\bigl\{S_c - S_p\bigr\}}{W} \right\rceil \times W,~~\forall c \in \mathbb{C}_p,
	\label{eq:obj}
\end{equation}

\noindent where $c$ is one of $p$'s consumers, denoted by $\mathbb{C}_p$. 
The ceiling operation is to enforce that a line buffer always stores multiples of a line and, thus, the actual line buffer size must be multiples of $W$, where $W$ is the image width.

%

%
%
%
%

\paragraph{Data Dependency.}
\Equ{eq:cc} states the data dependency requirement \boxed{R1}: an element must be in the line buffer before it can be read by its consumer(s).
Due to the nature of stencil computations, the data a consumer reads might span multiple image rows.
Therefore, a consumer must wait before the line buffer has certain number of pixels.
For instance, if we have a consumer whose stencil size is $3 \times 3$, the consumer must wait until the line buffer contains two full rows of pixels and one pixel from the third row (see \Fig{fig:lb1}).

Generally, for any producer-consumer pair $(p, c)$, the consumer start cycle must be delayed $(SH_c-1)\times W + 1$ cycles after the producer has started, where $SH_c$ is the height of the stencil window read by the consumer.

\Equ{eq:chw} states the on-chip memory constraint \boxed{R3}, which is far more complicated to express, which we discuss next.

\subsection{Modeling On-chip Memory Contention}
\label{sec:opt:hw}

\Equ{eq:chw} states that $B_{l,t}$, the number of accesses to any line $l$ in the line buffer at any given cycle $t$, must be no more than the number of ports ($P$) of the SRAM block that contains line $l$\footnote{In theory $P$ should be represented as $P_l$ to indicate that each SRAM could have a different number of ports. For simplicity purpose we assume that all the SRAMs in the hardware have the same number of ports. Our formulation, nevertheless, can be easily extended to support different port counts.}. The key is to mathematically express  $B_{l,t}$. To that end, we first express the set of lines that a stage accesses at each cycle.



Consider a pipeline stage $i$ accessing a line buffer at cycle $t$. The \textit{first} line accessed by stage $i$ at $t$, denoted by $L_{i, t}$, is:
\begin{equation}
	\label{eq:firstl}
	L_{i,t}=\left\lceil\frac{t-S_i}{W}\right\rceil.
\end{equation}

Thus, the \textit{Access Set} of stage $i$, i.e., the set of lines that stage $i$ accesses, at cycle $t$ is\footnote{A subtlety here is that the stencil height for the stage that writes to the line buffer is always 1.}:
\begin{equation}
	\label{eq:lines}
	\mathbb{A}_{i,t} = \left\{L_{i,t},~L_{i,t}+1,~\cdots,~L_{i,t}+SH_i-1\right\}.
\end{equation}


To satisfy the hardware constraint, no line can belong to the intersection of more than $P$ sets.
This is equivalent to saying that the intersection of any ($P+1$)-combination is always an empty set. Hence, the hardware constraint at each stage $i$ can be expressed mathematically as:

\begin{equation}
	\label{eq:hw1}
	\forall t~\forall \mathcal{T} \in \binom{\mathbb{N}_i}{P+1} ~~\bigcap_{i\in \mathcal{T}}\mathbb{A}_{i,t} = \emptyset,
\end{equation}

\noindent where $\binom{\mathbb{N}_i}{P+1}$ denotes the set of all ($P+1$)-combinations of $\mathbb{N}_i$, which itself is the set of all the stages that access the line buffer of stage $i$.



\begin{figure}[t]
    \centering
    \includegraphics[width=\linewidth]{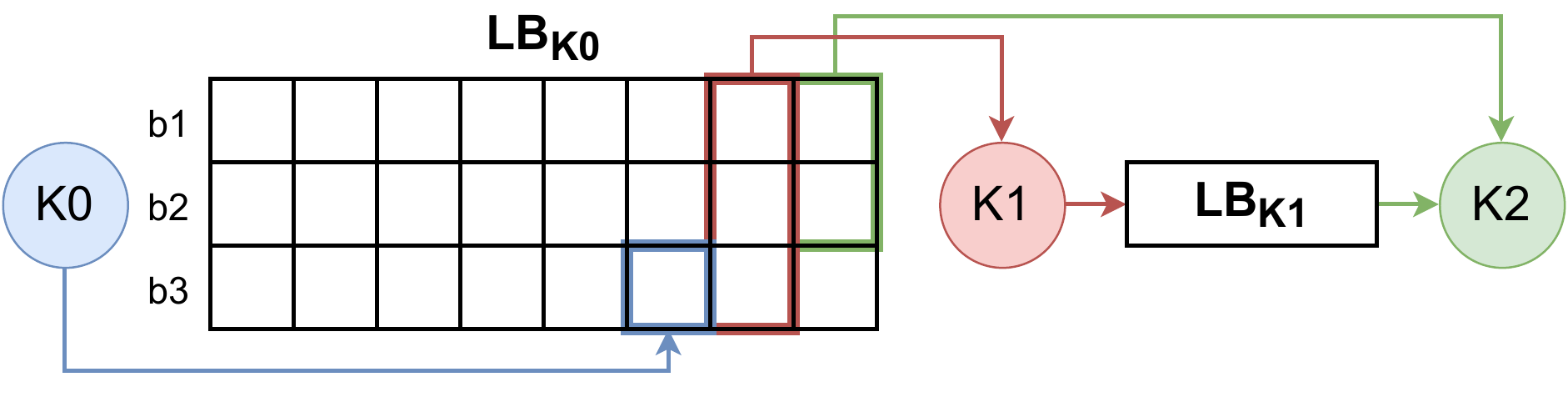}
    \caption{Example illustrating how to calculate Access Sets. At the current cycle $t$ shown in the figure, $K0$'s Access Set is $A_{0,t}$=(\textsf{b3}), $K1$'s Access Set is $A_{1,t}$=(\textsf{b1,b2,b3}), and $K2$'s Access Set is $A_{2,t}$=(\textsf{b1,b2}). Assuming each line stores in a dual-port SRAM, the hardware constraint would be $A_{0,t}\cap A_{1,t}\cap A_{2,t} = \emptyset$ (\Equ{eq:acs}), which after constraint pruning is reduced to $A_{0,t}\cap A_{2,t} = \emptyset$ (\Sect{sec:opt:cp}).}
    \label{fig:depth}
\end{figure}

To concretize \Equ{eq:hw1}, consider the example in \Fig{fig:depth}, where the line buffer $LB_{K0}$ is accessed by one producer ($K0$) and two consumers ($K1$ and $K2$). Assuming the SRAM has two ports, a common configuration in SRAM/BRAM blocks, the hardware constraint for $LB_{K0}$ is expressed as:
\begin{align}
  \forall t~~\mathbb{A}_{0,t}\cap \mathbb{A}_{1,t}\cap \mathbb{A}_{2,t}=\emptyset
  \label{eq:acs}
\end{align}

\noindent where $\mathbb{A}_{0, t}$, $\mathbb{A}_{1, t}$, and $\mathbb{A}_{2, t}$ are the Access Sets of stages $K0$, $K1$, and $K2$, respectively, at cycle $t$.

To enforce this constraint, one would enforce \textit{any} of the following three constraints, the intuition being if the intersection of any two sets is empty the intersection of all three sets is necessarily empty:
\begin{subequations}
\begin{align}
    & \forall t~~\mathbb{A}_{0,t}\cap \mathbb{A}_{1,t} = \emptyset, \label{eq:cond1} \\
    & \forall t~~\mathbb{A}_{0,t}\cap \mathbb{A}_{2,t} = \emptyset, \label{eq:cond2} \\
    & \forall t~~\mathbb{A}_{1,t}\cap \mathbb{A}_{2,t} = \emptyset. \label{eq:cond3}
\end{align}
\end{subequations}

\Equ{eq:cond1} -- \Equ{eq:cond3} involve calculating set intersections (or, equivalently, counting), not amenable to usual numerical optimizations. We must transform them to equivalent numerical expressions.

Without losing generality, consider the constraint $\forall t~~\mathbb{A}_{i,t}\cap \mathbb{A}_{j,t} = \emptyset$. Enforcing it is equivalent to enforcing that the last line written by $s_i$ (at any cycle $t$) must be above the first line read by $s_j$. That is:

\begin{equation}
    \forall t~~L_{i,t} + SH_i - 1 < L_{j, t},
	\label{eq:hw2}
\end{equation}

\noindent which, after applying \Equ{eq:firstl}, becomes:

\begin{equation}
	\label{eq:hw3}
		\forall t\: \left\lceil\frac{t-S_i}{W}\right\rceil+SH_i - 1 < \left\lceil\frac{t-S_j}{W}\right\rceil.
\end{equation}

\Equ{eq:hw3} depends on $t$, which does not have an upper bound.
Therefore, $t$ must be eliminated for the constraint to be usable. Our strategy is to (somehow) remove the ceiling operator ($\lceil~\rceil$), which would allow $t$ to be canceled out from both sides of \Equ{eq:hw3}. To that end, observe that:

\begin{align}
    x \leq \lceil x \rceil < x+1,
\end{align}

\noindent from which we can derive:

\begin{align}
    &\left\lceil\frac{t-S_i}{W}\right\rceil < \left(\frac{t-S_i}{W}\right) + 1,~~and \nonumber \\
    \left(\frac{t-S_j}{W}\right) \leq &\left\lceil\frac{t-S_j}{W}\right\rceil.
    \label{eq:hw4}
\end{align}

Combining \Equ{eq:hw4} with \Equ{eq:hw3}, we can transform \Equ{eq:hw3} into the following constraint:

\begin{align}
    & \forall t~~\left(\frac{t-S_i}{W}\right) + 1 + SH_i - 1 \leq \left(\frac{t-S_j}{W}\right) \nonumber \\
	\equiv~~& S_i - S_j \geq W\times SH_j.
    \label{eq:hw5}
\end{align}

\Equ{eq:hw5} is now independent of $t$. Note that \Equ{eq:hw5} is a stricter constraint than \Equ{eq:hw3}\footnote{The proof is a simple application of the transitivity of the ``less than'' ($<$) relation, which we omit here due to space limit.}, which means the solutions obtained with \Equ{eq:hw5} is a subset of those obtained with \Equ{eq:hw3}, sacrificing the solution optimality. The desirable trade-off, however, is that the constraint is independent of $t$. 
\Equ{eq:hw5} is then applied to re-write \Equ{eq:cond1}, \Equ{eq:cond2}, and \Equ{eq:cond3}.




\begin{figure*}[t]
    \centering
    \subfloat[Without line coalescing.]{
        \label{fig:lc1}
        \includegraphics[scale=0.42]{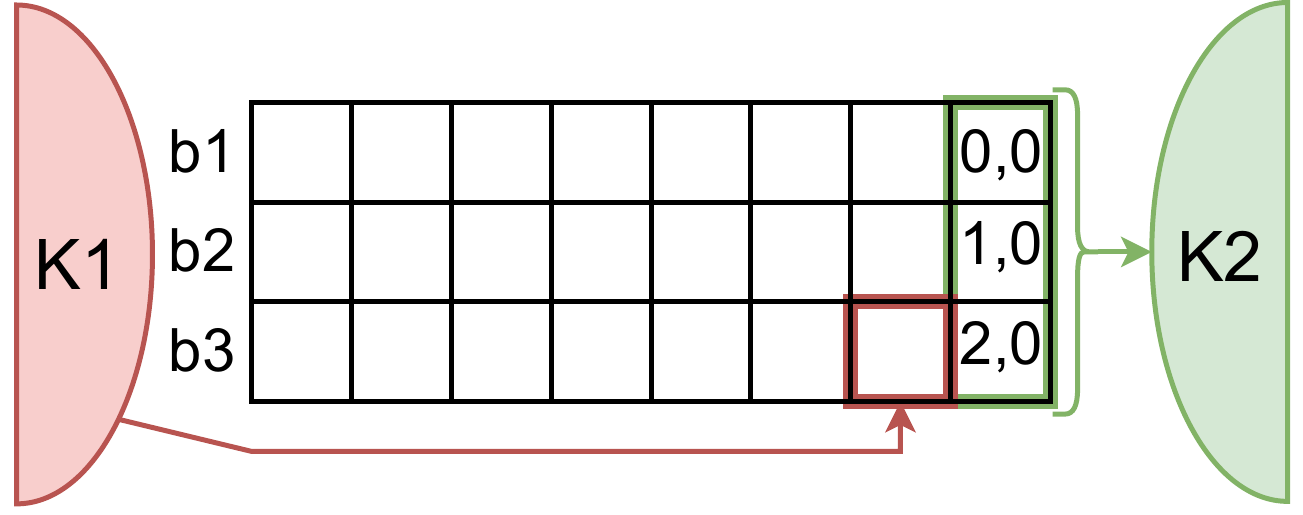}
    }
    \hspace{20pt}
    \subfloat[With line coalescing.]{
        \label{fig:lc2}
        \includegraphics[scale=0.42]{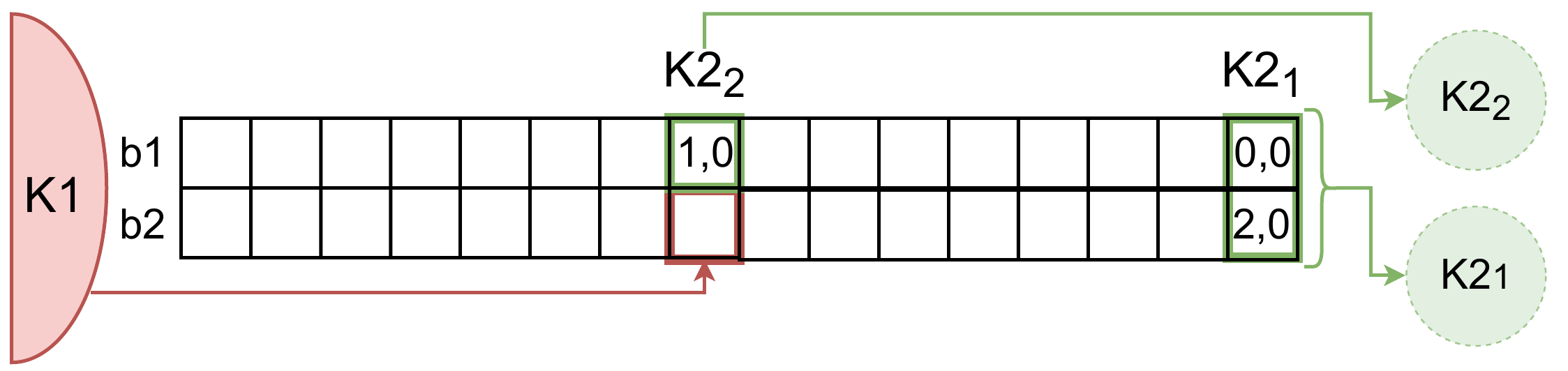}
    }
    \caption{Illustration of line coalescing optimization. Figure \protect\subref{fig:lc1} shows the implementation with line combination and \protect\subref{fig:lc2} shows that with line combination, assuming each memory block has two ports. In line combination, we place two consecutive lines in the same memory block. Effectively, the consumer stage $K2$ in the DAG is replaced with two ``virtual'' stages, $K2_1$ and $K2_2$, each of which has a stencil height of 2 and 1, respectively.}
    \label{fig:lc}
    \vspace{-10pt}
\end{figure*}

\subsection{Constraint Pruning}
\label{sec:opt:cp}

One potential issue is that the constraints in \Equ{eq:cond1}, \Equ{eq:cond2}, and \Equ{eq:cond3} are to be ``OR-ed'';
that is, only one of the three constraints needs to be satisfied.
Normally, this would require us to formulate three different sub-optimization problems, each of which considers one of the constraints individually. When a pipeline has multiple stages, each of which has constraints that are to be ``OR-ed'', the total number of sub-problems grows combinatorially.

To reduce the optimization time, we observe that constraints in \Equ{eq:cond1}, \Equ{eq:cond2}, and \Equ{eq:cond3} are not mutually exclusive, which allows us to prune some of them. We use the example in \Fig{fig:depth} to provide the intuition of constraint pruning, and then discuss how it is extended to general cases.

\paragraph{An Example.}
Observe that the constraint in \Equ{eq:cond2} is more \textit{relaxed} than that of \Equ{eq:cond1} and \Equ{eq:cond3}. That is, if \Equ{eq:cond1} (or \Equ{eq:cond3}) holds, \Equ{eq:cond2} necessarily holds:
\begin{subequations}
  \begin{align}
    & \forall t~~\mathbb{A}_{0,t}\cap \mathbb{A}_{1,t} = \emptyset \implies \forall t~~\mathbb{A}_{0,t}\cap \mathbb{A}_{2,t} = \emptyset, \label{eq:imp1} \\
    & \forall t~~\mathbb{A}_{1,t}\cap \mathbb{A}_{2,t} = \emptyset \implies \forall t~~\mathbb{A}_{0,t}\cap \mathbb{A}_{2,t} = \emptyset \label{eq:imp2}
  \end{align}
\end{subequations}

\noindent where $A \implies B$ reads ``$A$ implies $B$.''

Intuitively, \Equ{eq:imp1} holds because stage $K2$ data-depends on stage $K1$, which implies that $K2$ must start after $K1$. Therefore, at any time the first line $K2$ writes to must be below the first line $K1$ writes to (\Equ{eq:d1}), which in turn must be below the last line $K0$ writes to (\Equ{eq:d2}) given $\forall t~\mathbb{A}_{0,t} \cap \mathbb{A}_{1,t} = \emptyset$. Therefore, transitivity dictates that the first line $K2$ writes to is below the last line $K0$ writes to (\Equ{eq:d3}), hence $\forall t~\mathbb{A}_{0,t}\cap \mathbb{A}_{2,t} = \emptyset$.
\begin{subequations}
  \begin{align}
    & First(\mathbb{A}_{2,t}) > First(\mathbb{A}_{1,t}), \label{eq:d1} \\
    & First(\mathbb{A}_{1,t}) > Last(\mathbb{A}_{0,t}) \label{eq:d2}\\
    \implies & First(\mathbb{A}_{2,t}) > Last(\mathbb{A}_{0,t}) \label{eq:d3}
  \end{align}
\end{subequations}


The validity of \Equ{eq:imp2} can be similarly reasoned about using the fact that $K1$ data-depends on $K0$.

In general, given two constraints $A$ and $B$ that are to be ``OR-ed'', if $A$ is more relaxed than $B$, it is safe to eliminate $B$ without sacrificing optimality, because any solution that satisfies $B$ must also satisfy $A$. Therefore, in the example above we could eliminate constraints in \Equ{eq:cond1} and \Equ{eq:cond3}.


\paragraph{Generalization.} The example above shows that data dependency is key in eliminating redundant constraints.
In particular, data dependencies allow us to form partial orders $\preccurlyeq$ between stages. If there is a path in the DAG from stage $i$ to stage $j$, i.e., stage $j$ (directly or indirectly) depends on stage $i$, we have a partial order $i \preccurlyeq j$. Reflectivity holds for partial order relation: $i \preccurlyeq i$. We have the following theorem.

\underline{\textsc{Theorem}}. Given two generic constraints $C_1$ and $C_2$:
\begin{align}
  & C_1: \forall t~~\mathbb{A}_{x,t}\cap \mathbb{A}_{y,t} = \emptyset \\
  & C_2: \forall t~~\mathbb{A}_{z,t}\cap \mathbb{A}_{w,t} = \emptyset
  \label{eq:ec}
\end{align}

\noindent $C_1$ is more relaxed than $C_2$ (i.e., $C_2$ implies $C_1$) if $x \preccurlyeq z$, $w \preccurlyeq y$, and $SH_x \leq SH_z$.

\underline{\textsc{Proof}}. $x \preccurlyeq z$ leads to \Equ{eq:p1}, which combined with $SH_x \leq SH_z$ gives \Equ{eq:p2}; $w \preccurlyeq y$ gives \Equ{eq:p3}; $C_2$ gives \Equ{eq:p4}; transitivity thus yields \Equ{eq:p5}, which leads to $C_1$. Thus, $C_2$ implies $C_1$ and can be eliminated given $C_1$.
\begin{subequations}
  \begin{align}
    & First(\mathbb{A}_{x,t}) < First(\mathbb{A}_{z,t}), \label{eq:p1} \\
    & Last(\mathbb{A}_{x,t}) < Last(\mathbb{A}_{z,t}), \label{eq:p2} \\
    & First(\mathbb{A}_{w,t}) < First(\mathbb{A}_{y,t}), \label{eq:p3}\\
    & Last(\mathbb{A}_{z,t}) < First(\mathbb{A}_{w,t}), \label{eq:p4}\\
    \implies & Last(\mathbb{A}_{x,t}) < First(\mathbb{A}_{y,t}) \label{eq:p5}
  \end{align}
\end{subequations}

The theorem is a pruning rule: we examine each pair of constraints and eliminate the stricter one, if it exists, using the pruning rule. Note that given two constraints it is possible that one can \textit{not} make a judgment as to which is more relax/stricter, because there might not always be a partial order between two stages in a DAG.



%
%



\subsection{Problem Structure and Solver}
\label{sec:opt:solver}

The optimization problem we formulate is an Integer Linear Programming problem: the optimization variables are the start cycles of each stage---all integers; the objective function (\Equ{eq:objf}) and the constraints (\Equ{eq:cc} and \Equ{eq:hw5}) are all linear.
Note that the ceiling operations in the sub-terms of the objective function (\Equ{eq:obj}) can be removed without compromising the solution optimality, because minimizing $f(\lceil x \rceil)$ is equivalent to minimizing $f(x)$ given $f$ is monotonically increasingly
\footnote{More rigorously, we have: $\mathop{\mathrm{argmin}}f(x) \in \mathop{\mathrm{argmin}}f(\lceil x \rceil)$. That is, the solution that minimizes $f(x)$ is necessarily a solution that minimizes $f(\lceil x \rceil)$ for any monotonically increasing function $f$. To prove this, let $x_0 = \mathop{\mathrm{argmin}}f(x)$; then $\forall x~f(x) \ge f(x_0)$, so $\forall x~x \ge x_0$ (since $f$ is monotonically increasing). Thus, $\forall x~\lceil x \rceil \geq \lceil x_0 \rceil$, which means $\forall x~f(\lceil x \rceil) \geq f(\lceil x_0 \rceil)$, i.e., $x_0 \in \mathop{\mathrm{argmin}}f(\lceil x \rceil)$.}.
The ILP formulation let us use well-established solvers to quickly derive optimal line buffer designs.

\section{Line-Coalescing Optimization}
\label{sec:lc}

So far, we have assumed that each memory (e.g., SRAM/BRAM) block contains one line. It is possible, however, that the capacity of a memory block is large enough to hold multiple lines, in which case combining multiple lines into one single memory block would further reduce the memory requirement.
The challenge is how to generate the line-buffered pipeline under line coalescing. We show that our optimization formulation above can be naturally extended to support optimal line coalescing.


Consider the example in \Fig{fig:lc}, where there are two stages, a producer $K1$ and a consumer $K2$; the consumer operates on a stencil height of 3. Assume for now that each memory block has two ports.
\Fig{fig:lc1} and \Fig{fig:lc2} show the line buffer implementation without and with line coalescing, respectively.
Since each memory block has two ports, we could coalesce up to two lines into one memory block, which is shown in \Fig{fig:lc2}.
The three elements that $K2$ accesses are now spread across two, rather than three, memory blocks, as elements $(0,0)$ and $(1,0)$ are in the same memory block.

To express the line-coalesced pipeline to the optimizer, our observation is that line coalescing is equivalent to a transformation of the DAG, where the original stage $K2$ is replaced with two new ``virtual'' stages $K2_1$ and $K2_2$. In this example, \textsf{b1} is accessed simultaneously by the two virtual stages, whereas \textsf{b2} is accessed by only $K2_1$ (along with the producer $K1$). Thus, the virtual stage $K2_1$ now has an effective stencil height of 2, and the virtual stage $K2_2$ has an effective stencil height of 1.
Both virtual stages inherit the producer and consumers of the original stage $K2$.

We can generalize line coalescing to memory blocks with $P$ ports, where we can coalesce at most $P$ lines in one block and replace the 
original consumer stage with $P$ virtual stages. 
This transformation can be done offline, since it depends only on the algorithm DAG and stencil sizes. \Alg{alg:dag} describes the general transformation algorithm.

\RestyleAlgo{ruled}
\SetKwComment{Comment}{/* }{ */}
\SetKwRepeat{Do}{do}{while}

\begin{algorithm}
\caption{Line coalescing algorithm through DAG rewriting. Notation: $P$ is the number of ports, $SH_i$ is the stencil height read by stage $i$.}
\label{alg:dag}
\KwData{The original DAG}
\KwResult{The transformed DAG}
i = input node of the original DAG\;
 \While{i is not an empty node}{
  \If{i is not the input node}{
   $K$ = $\min(P,~SH_i)$\;
   split $i$ into $K$ virtual stages\;
   \For{each virtual stage $v$ split from $i$}{
    set $v$'s producer to $i$'s producer\;
    set $v$'s consumers to $i$'s consumers\;
   }
  }
  $i$ = next node through breath-first search\;
 }
 \Return the new DAG\;
\end{algorithm}

From the optimizer's perspective, the transformed DAG is nothing more than another pipeline except all the virtual stages belonging to the same physical stage must share the same start cycle, because logically they must act synchronously. Using the optimization formulation in \Equ{eq:opt}, the optimizer generates the optimal start cycles for every stage in the new DAG.
One special care the code generator takes is that virtual stages that belong to the same physical stage read from a different, but offline-determined offset. For instance in \Fig{fig:lc}, $K2_2$ will always read from an offset of $W$ (image width) from \textsf{b1}, where $K2_1$ and $K1$ have an offset of 0.

\paragraph{Remarks.} We note that the line coalescing optimization is fundamentally incompatible with the FIFO-based approach~\cite{chi2018soda} or designs that assume single-port memories~\cite{whatmough2019fixynn}---simply observe the data access behaviors in \Fig{fig:lc2}.

Line coalescing benefits both an FPGA and an ASIC backend. On FPGAs, BRAM block sizes are fixed on any particular board; forcing each block to hold only one line could result in internal fragmentation of BRAM blocks. ASICs designers could customize the memory for an algorithm; they could properly size the memory blocks to permit line coalescing to reduce the overall area (\Sect{sec:res:dse}).

\section{Experimental Methodology}
\label{sec:exp}

\paragraph{Compiler Implementation.} We implement our compiler in Python with about 1,500 lines of code.
We use Google's optimization library ``or-tools''~\cite{ortools} for solving the ILP problem.

\paragraph{Hardware Platform.}
We evaluate both an ASIC flow and an FPGA flow.
We use a Xilinx Spartan-7 FPGA board (xa7s100fgga488-2I) for evaluation. The board has 120 BRAM blocks and each block is of size 36 Kbits. Each block can be configured as either a single-port or a dual-port memory block.
We assume SRAM blocks are available at 64 KB for line buffers in the ASIC backend. We evaluate two image resolutions: $320p$ ($480 \times 320$) and $1080p$ ($1920 \times 1080$). The SRAM and BRAM block sizes make sure line coalescing applies to $320p$ but not $1080p$, since the block size is not large enough to hold multiple rows in an $1080p$ image.

For the FPGA backend the generated Verilog code goes through the FPGA synthesis and layout flow using Vivado Design Suite 2021.1. We use Vivado's resource monitor to report the BRAM usage.
\hl{The FPGA communicates with the host through AXI DMA. Through DMA, we first load the input image to the BRAM from the host memory. The frame rate reported (1 pixel per cycle) is the throughput after the accelerator has started, i.e., steady-state throughput.}
For each design, we perform post-implementation functional simulation to obtain the switching activity, which is then used by Vivado's power analysis tool to obtain power consumption.

For the ASIC backend we build a cycle-level simulator to simulate the line-buffered pipelines. We use 
the open-source memory compiler OpenRAM\cite{openram} with FreePDK45~\cite{freepdk45} to estimate the per-access SRAM power, which is then combined with the number of accesses given by our simulator to estimate the total memory power.

Since the goal of this paper is to reduce on-chip memory size and energy, we primarily report results related to the on-chip memory \hl{but will also show savings for the entire accelerator. Note that the memory area dominates the accelerator area, so the memory area/power savings are expected to translate to similar total area/power savings. In the ASIC backend, the SRAM area contributes to, on average, 79.8\% and 92.7\% of the total accelerator area across all algorithms on $320p$ and $1080p$ images, respectively. The reason memory area dominates is that there are very few PEs in line-buffered accelerators: to execute a $3 \times 3$ convolution, regardless of the input, we require only $3 \times 3$ MAC units (see \mbox{\Fig{fig:line_buffer}}). The total area, on average, is 0.65 $mm^2$ and 1.84 $mm^2$, for the two resolutions, respectively, and the total average power is 72.9 $mW$ and 98.3 $mW$, for the two resolutions, respectively.}

\paragraph{Algorithms.}
We evaluate common image processing algorithms listed in \Tbl{tab:algos}, where each algorithm either ends with an ``\textsf{-s}'', indicating it has only single-consumer stages or with an ``\textsf{-m}'', indicating it has at least one multiple-consumer stage. Both \textsf{Canny} and \textsf{Harris} has two versions depending on the implementation details.


\begin{table}[t]
\caption{Evaluation algorithms. \textsf{-s} or \textsf{-m} indicates if an algorithm has only single-consumer stages or has at least one multiple-consumer stage, respectively. The last column shows the number of multiple-consumer (MC) stages.}
\label{tab:algos}
\resizebox{\columnwidth}{!}{
\renewcommand*{\arraystretch}{1}
\renewcommand*{\tabcolsep}{4pt}
\begin{tabular}{llcc}
\toprule[0.15em]
\textbf{Algorithm} & \textbf{Description}                     & \textbf{\# Stages} & \textbf{\# of MC Stages} \\
\midrule[0.05em]
Canny-s            & \multirow{2}{*}{Canny edge detection}    & 9                  & 0                                                                                        \\
Canny-m            &                                          & 10                 & 1                                                                                        \\
Harris-s           & \multirow{2}{*}{Harris corner detection} & 7                  & 0                                                                                        \\
Harris-m           &                                          & 7                  & 1                                                                                        \\
Unsharp-m          & Unsharp masking                          & 5                  & 1                                                                                        \\
Xcorr-m            & Cross correlation                        & 3                  & 1                                                                                        \\
Denoise-m          & Image denoise                            & 5                  & 2                                                                                        \\ \bottomrule[0.15em]
\end{tabular}
}
\vspace{-10pt}
\end{table}


\paragraph{Baselines and Variants.} We compare with three common line-buffered image processing accelerators.

\begin{itemize}
	\item FixyNN~\cite{whatmough2019fixynn}, which is based on the same design described in \Sect{sec:background} but uses only single-port SRAMs.
	\item SODA~\cite{chi2018soda}, which uses FIFO to implement line buffers and splits FIFOs to support multiple-consumer stages (\Sect{sec:motivation}). The FIFOs are implemented using dual-port SRAMs (rather than shift registers).
	\item Darkroom~\cite{hegarty2014darkroom}, which linearizes multiple-consumer pipelines (\Sect{sec:motivation}) and uses two-port SRAMs.
\end{itemize}



We consider two variants of our framework: \mode{Ours} and \mode{Ours+LC}. The latter adds line-coalescing to the former.
\section{Evaluation Results}
\label{sec:res}

We first show that our compiler maintains the theoretical maximum throughput (\Sect{sec:res:perf}) and is fast to execute (\Sect{sec:res:comp}). We then show that the hardware generated by our framework consumes less memory resource (\Sect{sec:res:mem}) and lower power (\Sect{sec:res:power}) compared to existing methods. Finally, we show that our framework can help customize memory modules for individual algorithms (\Sect{sec:res:dse}).

\subsection{Throughput and Latency}
\label{sec:res:perf}

Across all algorithms, hardware generated by our compiler maintains a constant throughput of one pixel per cycle, the target laid out and justified in \Sect{sec:opt:intuition}.
\mode{Ours} increases the average end-to-end latency by only 0.01\% over Darkroom and SODA. Thus, the memory and power savings shown later come with no speed degradation.

\subsection{Compilation Speed}
\label{sec:res:comp}



On average, our compiler takes 14.5 ms to generate the Verilog code across all the algorithms. For multiple-consumer algorithms, constraint pruning (\Sect{sec:opt:cp}) speeds up the compilation time by 4$\times$ on average. This is achieved by pruning redundant constraints that would have led to many sub-optimization problems. Compared to Darkroom's linearization compiler, our compiler, on an average, compiles 37.4\% faster. This is because linearization adds adds dummy stages, which adds more constraints to the ILP.

\paragraph{Scalability.} We also sweep across different pipelines of varying length from 9 to 60. In each algorithm a third of the stages had multiple consumers. It took 8.7 ms for our compiler to compile 9 stage pipeline and 8.1 s to compile the 60 stage pipeline, showing the scalability of our compiler.

\begin{figure}
  \vspace{2pt}
  \centering
  \subfloat[SRAM size.]{
        \label{fig:bram320}
        \includegraphics[width=\columnwidth]{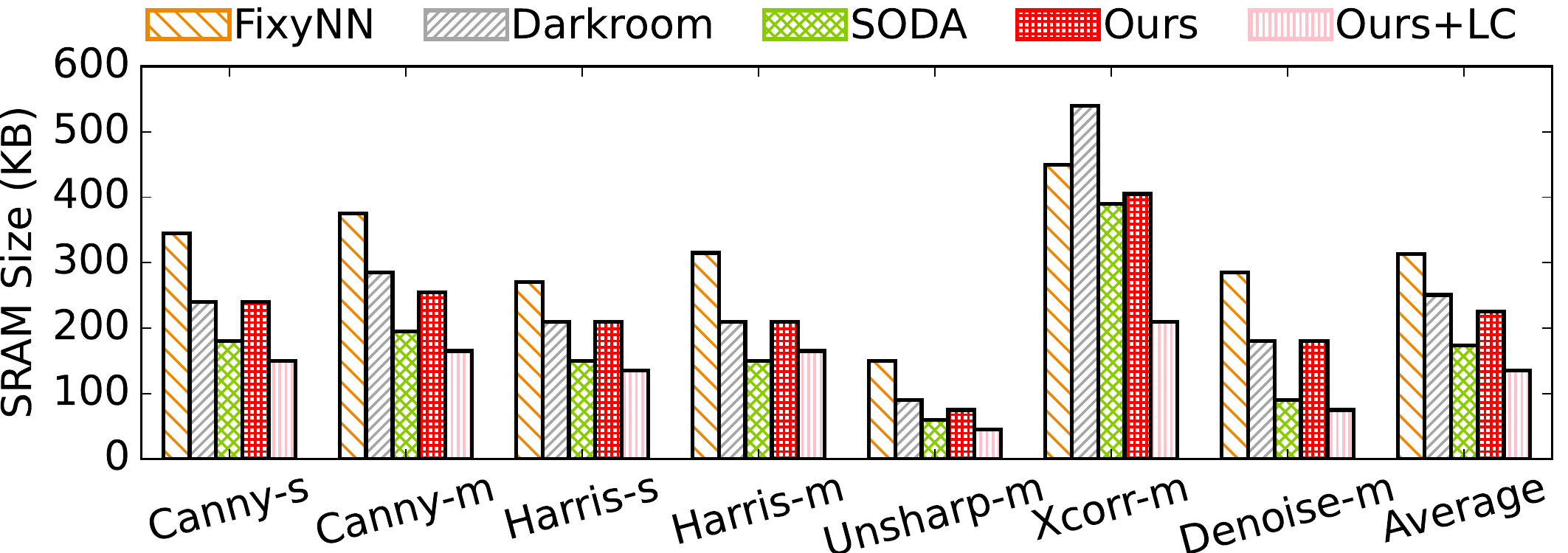}
  }\\
  \subfloat[Power consumption.]{
        \label{fig:power320}
        \includegraphics[width=\columnwidth]{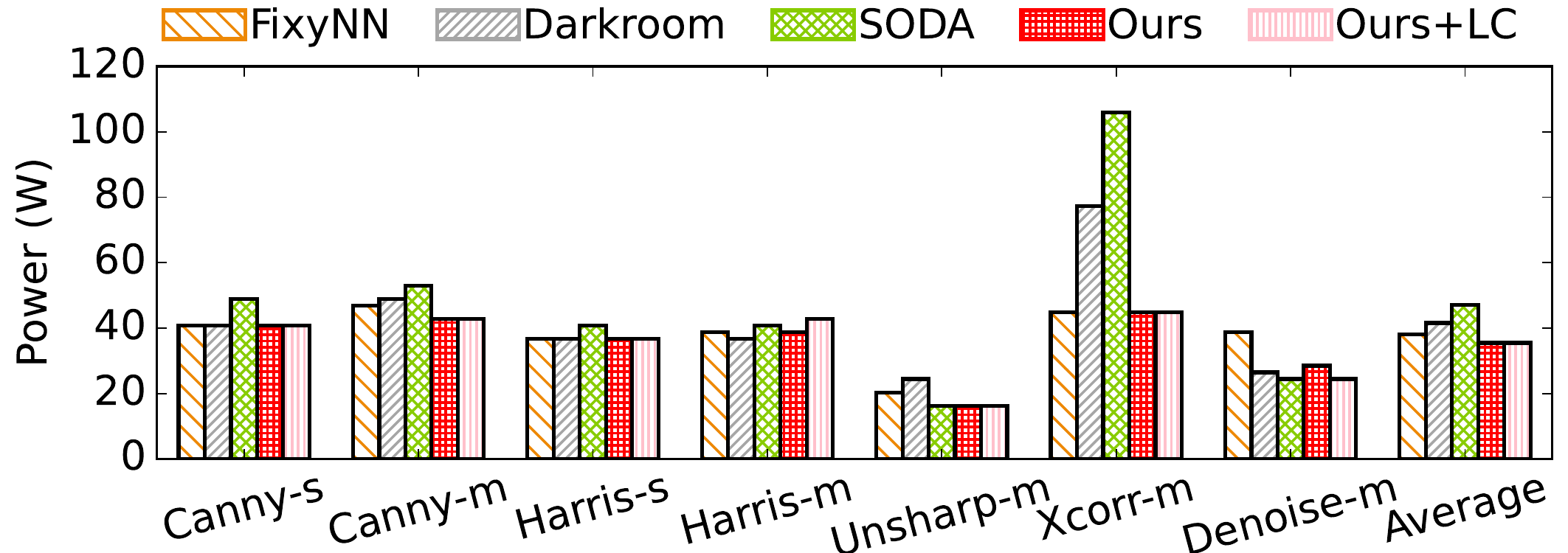}
  }
  \caption{SRAM and power comparison on $320p$ images.}
\end{figure}

\begin{figure}
  \centering
  \subfloat[SRAM size.]{
        \label{fig:bram1080}
        \includegraphics[width=\columnwidth]{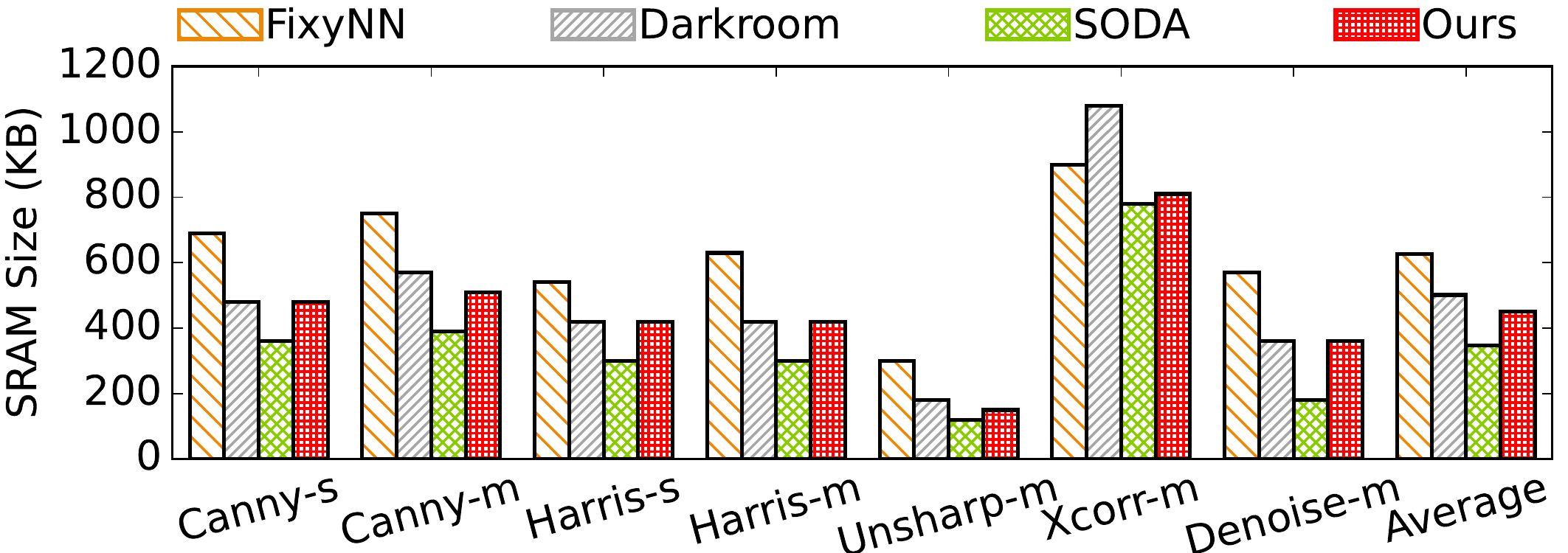}
  }\\
  \subfloat[Power consumption.]{
        \label{fig:power1080}
        \includegraphics[width=\columnwidth]{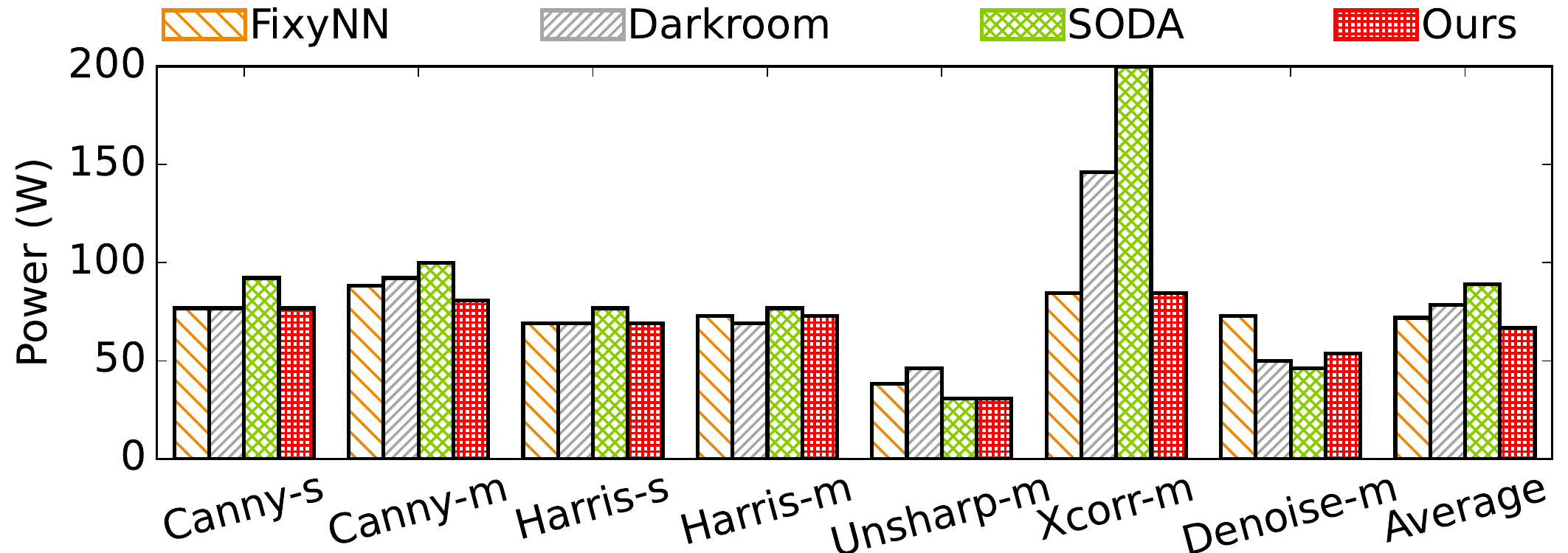}
  }
  \caption{SRAM and power consumption comparison on $1080p$ images. Note that \mode{Ours+LC} is not shown, since the SRAM size is not large enough to hold more than one lines.}
  \vspace{-10pt}
\end{figure}

\subsection{On-Chip Memory Requirement Reduction}
\label{sec:res:mem}



Accelerators generated by our framework reduce the on-chip memory size significantly. \Fig{fig:bram320} compares the SRAM size of the hardware generated by our framework and the three baselines on $320p$ images. Averaging over all the algorithms, \mode{Ours} reduces the SRAM size by 28.0\% and 10.2\% compared to FixyNN and Darkroom, respectively. After the line-coalescing optimization is applied, the SRAM savings over the baselines increase to 86.0\% and 56.8\%, respectively.

The SRAM saving on multiple-consumer algorithms is noticeably higher than that on single-consumer algorithms, highlighting the benefits of our framework on the former.
On multiple-consumer algorithms, algorithm linearization adds dummy stages and increases line buffer size.
FixyNN always has a higher SRAM requirement than \mode{Ours}, even on single-consumer algorithms, because it uses only single-port memory blocks, where no two stages are allowed to overlap.
The SRAM saving is particularly significant on \textsf{Xcorr-m}. This is because when linearizing \textsf{Xcorr-m}, one of the stages that are replicated operates on a tall stencil window (18$\times$1); replicating that stage adds a lot of additional SRAM blocks.

The SRAM requirement of \mode{Ours} is 31.0\% higher than SODA, because SODA, being a FIFO-based approach, is able to implement the last line in the line buffer (the line being written to by the producer) as DFFs (\Fig{fig:fifo1}).
With line coalescing, \mode{Ours+LC} reduces the SRAM requirement by 28.5\% compared to SODA.

\Fig{fig:bram1080} shows that the SRAM saving trend on $1080p$ inputs is similar to that on $320p$ inputs, except that the line coalescing optimization could not be applied to $1080p$ images, since the SRAM block size is not large enough to hold more than one line, as discussed in \Sect{sec:exp}.

\hl{\mbox{\paragraph{Accelerator Results.}} Memory area dominates the accelerator area, as discussed in \mbox{\Sect{sec:exp}}. Thus, the memory size saving translates to similar total accelerator area saving. For instance, compared to FixyNN and Darkroom on $320p$ images, \mbox{\mode{Ours+LC}} saving the total area by 51.2\% and 41.9\%, respectively. The savings are 27.9\% and 12.9\% on $1080p$ images.}

\paragraph{FPGA Results.} Due to the space limit we summarize the main results from the FPGA implementation. On $1080p$ images, \mode{Ours} reduces the BRAM size by 28.1\% and 10.2\% compared to FixyNN and Darkroom, respectively, and increase the BRAM usage by 22.8\% over SODA, for the same reason described above. On our FPGA, \mode{Ours} uses 37.5\% of the BRAM blocks as opposed to 41.8\% by Darkroom.

\paragraph{Multiple Algorithms.} Our goal is \textit{not} a generic stencil accelerator that runs multiple algorithms. Rather, we focus on accelerators that are specialized for a given algorithm.
Nevertheless, by reducing memory usage our compiler can also help generic stencil accelerators that has one single memory system --- by accommodating more algorithms simultaneously. For instance, on our FPGA with 120 BRAM blocks, FixyNN and Darkroom could not simultaneously execute all six algorithms even in the $320p$ resolution because of the BRAM constraint. With \mode{Ours+LC}, however, the FPGA can accommodate all six algorithms using only 84 BRAM blocks.

\subsection{Power Consumption Reduction}
\label{sec:res:power}

We also generate accelerators that consume lower memory power compared to all baselines. \Fig{fig:power320} compares the power consumption on $320p$ images.
On average, \mode{Ours} consumes 7.8\%, 13.8\%, and 56.0\% less power than FixyNN, Darkroom, and SODA respectively. 
Line coalescing does not change the power by much, since the total memory accesses remain roughly the same.
The power savings over Darkroom and FixyNN come from the SRAM size reduction.
For instance, while FixyNN, which uses only single-port memories, has lower per-access power, using single-port memories results in more SRAMs, increasing the total power.

It is interesting to observe that \mode{Ours} has lower power compared to SODA even though \mode{Ours} require more SRAM arrays than SODA (\Fig{fig:bram320}).
This is because SODA uses FIFOs, which have to serve two accesses \textit{every} cycle. In our design, all but one SRAM array serve only one access per cycle, leading to an overall power reduction.
The power saving of \mode{Ours+LC} over SODA comes from both reducing the SRAM requirement and avoiding power-hungry FIFOs.

\Fig{fig:power1080} compares the power consumption using $1080p$ images.
\mode{Ours} consumes 7.8\%, 13.8\%, and 56.0\% less power than FixyNN, Darkroom, and SODA, respectively. Again, even though \mode{Ours} uses more SRAM than SODA, it has lower power consumption because it avoids power-hungry FIFOs.

\hl{\mbox{\paragraph{Accelerator Results.}} Memory power savings translate to similar accelerator-level savings. On $320p$ images \mbox{\mode{Ours}} consumes 11.7\% and 15.2\% less power compared to Darkroom and SODA, respectively; on $1080p$ images the savings are 11.9\% and 18.3\%.}

\paragraph{FPGA Results.} The power saving trend on the FPGA is similar. On $1080p$ inputs, \mode{Ours} consumes 19.7\%, 5.8\%, and 17.7\% less power than FixyNN, Darkroom, and SODA, respectively. The FPGA power saving is lower, because FPGAs consume non-trivial static power.

\subsection{Design Space Exploration}
\label{sec:res:dse}

\begin{figure}
    \vspace{-8pt}
    \subfloat[\textsf{Canny-m}.]{
        \label{fig:dse11}
        \includegraphics[width=0.48\linewidth]{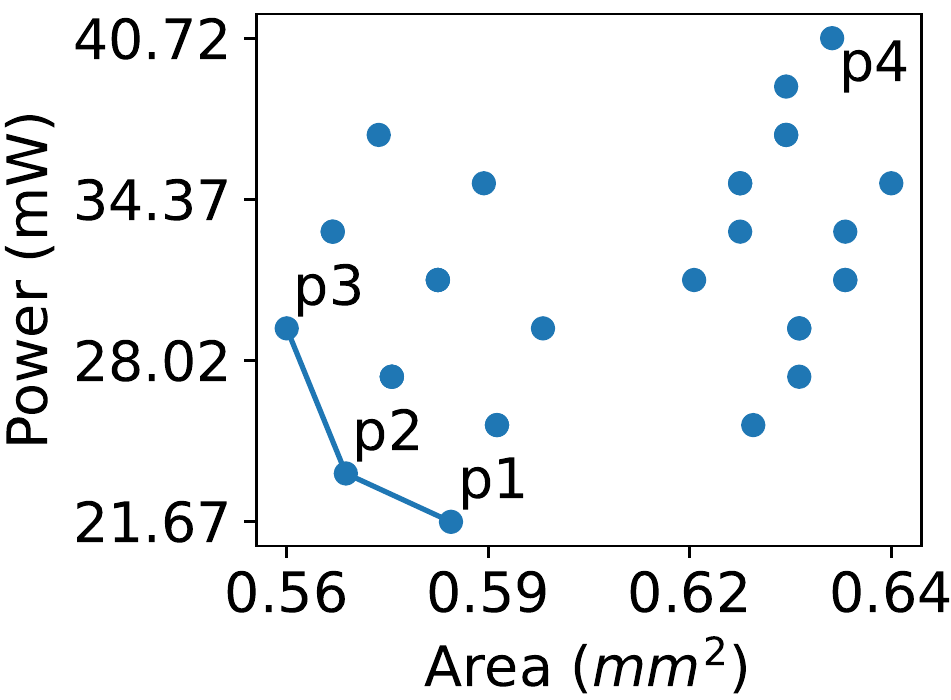}
    }
    \subfloat[\textsf{Denoise-m}.]{
        \label{fig:dse12}
        \includegraphics[width=0.48\linewidth]{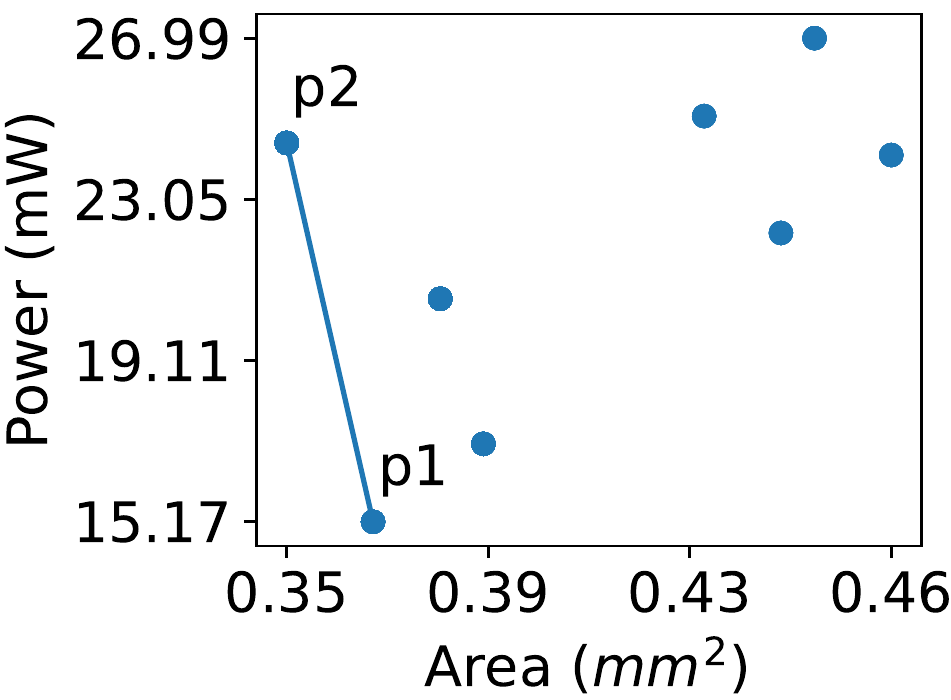}
    }
    \caption{Representative power-vs-area trade-offs under $320p$ for \textsf{Canny-m} and \textsf{Denoise-m}. For each algorithm, we sweep the memory configuration for each stage and generate a corresponding optimal design. Each stage is allowed to use either  double-port memory (DP) or DP with line coalescing (DPLC). The Pareto-optimal designs vary with algorithms.}
    \vspace{-10pt}
    \label{fig:dsedlc}
\end{figure}



Evaluation so far assumes that one type of line buffer is used for all the stages in all algorithms, which is the only option in prior work~\cite{whatmough2019fixynn, hegarty2014darkroom, chi2018soda}.
Since our framework permits specifying arbitrary memory configurations,
it can be used (by ASIC designers) to create custom memory modules to explore the power-vs-area trade-off in an \textit{algorithm-specific} manner.
Specifically, we allow each line buffer in an algorithm to be implemented as either a 
double-port memory (DP) or DP with line coalescing (DPLC).
For each algorithm, we then sweep all the possible memory configuration combinations and generate the corresponding optimal design for each combination.
For example, if there are four stages in an algorithm, we would end up with 16 different designs.

We observe that the Pareto-optimal designs vary with algorithm.
Using $320p$ as an example,  \Fig{fig:dse11} shows the power-vs-area comparison for \textsf{Canny-m}, where there are three Pareto-optimal designs.
P1 uses the DP configuration for all the stages; in P2, one stage uses the DPLC configuration, and in P3 two stages use the DPLC configuration. 
In contrast, \Fig{fig:dse12} shows another trade-off pattern possessed by \textsf{Denoise-m}, where there are only two Pareto optimal configurations. In this case, P1 uses only DP for all the stages and P2 uses only DPLC for all the stages.

We particularly note that for \textsf{Canny-m} the design that uses DPLC for all stages is P4 in \Fig{fig:dse11}, which is far worse than the three Pareto-optimal designs of \textsf{Canny-m}.
DPLC reduces the number of SRAM arrays and the total area, but the per access power also increases.
Thus, the total power depends on the total memory accesses, which is necessarily algorithm-specific, an exploration that is uniquely enabled by our tool.

\section{Related Work}
\label{sec:rw}

Agile accelerator design has received considerable attention. A recent theme is languages that close the gap between high-level algorithm semantics and hardware design~\cite{koeplinger2018spatial,lai2019heterocl,stewart2018ripl,nigam2020predictable}. These languages allow high-level descriptions of an algorithm and expose the hardware as a set of primitive components. Our work focuses on one particular kind of algorithm domain (image processing), and one particular aspect of hardware components: line-buffered on-chip memory.

Dahlia~\cite{nigam2020predictable} uses a type system to reject programs that could have unpredictable behaviors in hardware --- memory contention being one of them. Our work encodes memory-port contention as a constraint and generates (resource-optimal) hardware that avoids contention.
Aetherling~\cite{durst2020type} is a DSL that compiles high-level image processing algorithms to hardware with the focus of exploring resource-vs-throughput trade-offs. It does not guarantee minimizing memory resource consumption, which we do.
HeteroHalide~\cite{li2020heterohalide} and HeteroCL~\cite{lai2019heterocl} synthesize hardware accelerators and rely on SODA~\cite{chi2018soda} to generate the on-chip memory system.
HalideHLS~\cite{pu2017programming} generates accelerators for image processing algorithms, but rely on the user to optimize the on-chip memory.
DSAGEN~\cite{weng2020dsagen} annotates algorithms using pragmas and automatically searches a large architecture design space for a range of algorithms.

An orthogonal effort is mapping/scheduling an algorithm onto a fixed hardware substrate. Prior work uses constrained optimization methods~\cite{nowatzki2013general, huang2021cosa,feng2019asv}, targeting mostly deep learing workloads. They leverage the fact that the accelerator design space can usually be parameterized and behaviors of algorithms of interest can be mechanically modeled, two traits that our work leverages, too.

\section{Conclusion and Future Work}
\label{sec:conc}

This paper presents a framework that automatically synthesizes accelerators for image processing.
The key is an optimization formulation that permits expressing memory contention as a generic constraint. The explicit memory-constrained optimization allows us to avoid manual heuristics and customize designs in an application-specific way. We show that accelerators generated by our framework reduce on-chip memory usage and power by up to 86.0\% and 62.9\%, respectively, when compared to state-of-the-art methods.

We demonstrate our framework on image processing because it is central to emerging applications such as autonomous machines. 
Fundamentally, however, our framework is not limited to image processing; rather, it generalizes to all stencil algorithms, which are central to scientific computing, many of which operate on generic meshes rather than images~\mbox{\cite{holewinski2012high}}.
Our main technical novelty, i.e., expressing on-chip memory contention in a way that is amenable to numerical optimization, generalizes to any regular algorithm accessing arbitrary on-chip memories, not just line buffers.

Interesting lines of future work include automatically synthesizing sparse image processing accelerators~\cite{guo2020accelerating, liu2022s2ta} and accelerators that operate on irregular visual data such as meshes and point clouds~\cite{xu2019tigris, feng2020mesorasi, feng2022crescent}. These are application domains where accelerators are almost exclusively manually designed.
\section{Acknowledgements}

We thank the anonymous reviewers from HPCA 2023 and ISCA 2023 for their valuable feedback. Jingwen Leng and Yuhao Zhu are the corresponding co-authors. The work was supported, in part, by NSF under grants \#2044963 and \#2126642.

\bibliographystyle{ACM-Reference-Format}
\balance
\bibliography{references}


\begin{thebibliography}{40}


\ifx \showCODEN    \undefined \def \showCODEN     #1{\unskip}     \fi
\ifx \showDOI      \undefined \def \showDOI       #1{#1}\fi
\ifx \showISBNx    \undefined \def \showISBNx     #1{\unskip}     \fi
\ifx \showISBNxiii \undefined \def \showISBNxiii  #1{\unskip}     \fi
\ifx \showISSN     \undefined \def \showISSN      #1{\unskip}     \fi
\ifx \showLCCN     \undefined \def \showLCCN      #1{\unskip}     \fi
\ifx \shownote     \undefined \def \shownote      #1{#1}          \fi
\ifx \showarticletitle \undefined \def \showarticletitle #1{#1}   \fi
\ifx \showURL      \undefined \def \showURL       {\relax}        \fi
\providecommand\bibfield[2]{#2}
\providecommand\bibinfo[2]{#2}
\providecommand\natexlab[1]{#1}
\providecommand\showeprint[2][]{arXiv:#2}

\bibitem[fre({[n.\,d.]})]%
        {freepdk45}
 \bibinfo{year}{[n.\,d.]}\natexlab{}.
\newblock \bibinfo{title}{{FreePDK45}}.
\newblock
  \bibinfo{howpublished}{\url{https://eda.ncsu.edu/freepdk/freepdk45/}}.
\newblock


\bibitem[ort({[n.\,d.]})]%
        {ortools}
 \bibinfo{year}{[n.\,d.]}\natexlab{}.
\newblock \bibinfo{title}{{Google OR-Tools}}.
\newblock
  \bibinfo{howpublished}{\url{https://developers.google.com/optimization}}.
\newblock


\bibitem[mal({[n.\,d.]})]%
        {malic55}
 \bibinfo{year}{[n.\,d.]}\natexlab{}.
\newblock \bibinfo{title}{{Mali-C55}}.
\newblock
  \bibinfo{howpublished}{\url{https://developer.arm.com/Processors/Mali-C55}}.
\newblock


\bibitem[qcs({[n.\,d.]})]%
        {qcspectra}
 \bibinfo{year}{[n.\,d.]}\natexlab{}.
\newblock \bibinfo{title}{{Snapdragon Makes Significant Leap for Mobile Cameras
  with Qualcomm Spectra Image Signal Processor and Snapdragon Sight}}.
\newblock
  \bibinfo{howpublished}{\url{https://futurumresearch.com/snapdragon-makes-significant-leap-for-mobile-cameras-with-qualcomm-spectra-image-signal-processor-and-snapdragon-sight/}}.
\newblock


\bibitem[Bagni et~al\mbox{.}(2017)]%
        {bagni2017demystifying}
\bibfield{author}{\bibinfo{person}{Daniele Bagni}, \bibinfo{person}{Pari
  Kannan}, {and} \bibinfo{person}{Stephen Neuendorffer}.}
  \bibinfo{year}{2017}\natexlab{}.
\newblock \showarticletitle{Demystifying the Lucas-Kanade optical flow
  algorithm with Vivado HLS}.
\newblock \bibinfo{journal}{\emph{Tech. note XAPP1300. Xilinx}}
  (\bibinfo{year}{2017}).
\newblock


\bibitem[Chandramoorthy et~al\mbox{.}(2015)]%
        {ivs}
\bibfield{author}{\bibinfo{person}{Nanchini Chandramoorthy},
  \bibinfo{person}{Giuseppe Tagliavini}, \bibinfo{person}{Kevin Irick},
  \bibinfo{person}{Antonio Pullini}, \bibinfo{person}{Siddharth Advani},
  \bibinfo{person}{Sulaiman Al~Habsi}, \bibinfo{person}{Matthew Cotter},
  \bibinfo{person}{John Sampson}, \bibinfo{person}{Vijaykrishnan Narayanan},
  {and} \bibinfo{person}{Luca Benini}.} \bibinfo{year}{2015}\natexlab{}.
\newblock \showarticletitle{{Exploring Architectural Heterogeneity in
  Intelligent Vision Systems}}. In \bibinfo{booktitle}{\emph{Proc. of HPCA}}.
\newblock


\bibitem[Chi et~al\mbox{.}(2018)]%
        {chi2018soda}
\bibfield{author}{\bibinfo{person}{Yuze Chi}, \bibinfo{person}{Jason Cong},
  \bibinfo{person}{Peng Wei}, {and} \bibinfo{person}{Peipei Zhou}.}
  \bibinfo{year}{2018}\natexlab{}.
\newblock \showarticletitle{SODA: Stencil with optimized dataflow
  architecture}. In \bibinfo{booktitle}{\emph{2018 IEEE/ACM International
  Conference on Computer-Aided Design (ICCAD)}}. IEEE, \bibinfo{pages}{1--8}.
\newblock


\bibitem[Durst et~al\mbox{.}(2020)]%
        {durst2020type}
\bibfield{author}{\bibinfo{person}{David Durst}, \bibinfo{person}{Matthew
  Feldman}, \bibinfo{person}{Dillon Huff}, \bibinfo{person}{David Akeley},
  \bibinfo{person}{Ross Daly}, \bibinfo{person}{Gilbert~Louis Bernstein},
  \bibinfo{person}{Marco Patrignani}, \bibinfo{person}{Kayvon Fatahalian},
  {and} \bibinfo{person}{Pat Hanrahan}.} \bibinfo{year}{2020}\natexlab{}.
\newblock \showarticletitle{Type-directed scheduling of streaming
  accelerators}. In \bibinfo{booktitle}{\emph{Proceedings of the 41st ACM
  SIGPLAN Conference on Programming Language Design and Implementation}}.
  \bibinfo{pages}{408--422}.
\newblock


\bibitem[Feng et~al\mbox{.}(2022)]%
        {feng2022crescent}
\bibfield{author}{\bibinfo{person}{Yu Feng}, \bibinfo{person}{Gunnar Hammonds},
  \bibinfo{person}{Yiming Gan}, {and} \bibinfo{person}{Yuhao Zhu}.}
  \bibinfo{year}{2022}\natexlab{}.
\newblock \showarticletitle{Crescent: taming memory irregularities for
  accelerating deep point cloud analytics}. In
  \bibinfo{booktitle}{\emph{Proceedings of the 49th Annual International
  Symposium on Computer Architecture}}. \bibinfo{pages}{962--977}.
\newblock


\bibitem[Feng et~al\mbox{.}(2020)]%
        {feng2020mesorasi}
\bibfield{author}{\bibinfo{person}{Yu Feng}, \bibinfo{person}{Boyuan Tian},
  \bibinfo{person}{Tiancheng Xu}, \bibinfo{person}{Paul Whatmough}, {and}
  \bibinfo{person}{Yuhao Zhu}.} \bibinfo{year}{2020}\natexlab{}.
\newblock \showarticletitle{Mesorasi: Architecture support for point cloud
  analytics via delayed-aggregation}. In \bibinfo{booktitle}{\emph{2020 53rd
  Annual IEEE/ACM International Symposium on Microarchitecture (MICRO)}}. IEEE,
  \bibinfo{pages}{1037--1050}.
\newblock


\bibitem[Feng et~al\mbox{.}(2019)]%
        {feng2019asv}
\bibfield{author}{\bibinfo{person}{Yu Feng}, \bibinfo{person}{Paul Whatmough},
  {and} \bibinfo{person}{Yuhao Zhu}.} \bibinfo{year}{2019}\natexlab{}.
\newblock \showarticletitle{Asv: Accelerated stereo vision system}. In
  \bibinfo{booktitle}{\emph{Proceedings of the 52nd Annual IEEE/ACM
  International Symposium on Microarchitecture}}. \bibinfo{pages}{643--656}.
\newblock


\bibitem[Gan et~al\mbox{.}(2021)]%
        {gan2021eudoxus}
\bibfield{author}{\bibinfo{person}{Yiming Gan}, \bibinfo{person}{Yu Bo},
  \bibinfo{person}{Boyuan Tian}, \bibinfo{person}{Leimeng Xu},
  \bibinfo{person}{Wei Hu}, \bibinfo{person}{Shaoshan Liu},
  \bibinfo{person}{Qiang Liu}, \bibinfo{person}{Yanjun Zhang},
  \bibinfo{person}{Jie Tang}, {and} \bibinfo{person}{Yuhao Zhu}.}
  \bibinfo{year}{2021}\natexlab{}.
\newblock \showarticletitle{Eudoxus: Characterizing and accelerating
  localization in autonomous machines industry track paper}. In
  \bibinfo{booktitle}{\emph{2021 IEEE International Symposium on
  High-Performance Computer Architecture (HPCA)}}. IEEE,
  \bibinfo{pages}{827--840}.
\newblock


\bibitem[Guo et~al\mbox{.}(2020)]%
        {guo2020accelerating}
\bibfield{author}{\bibinfo{person}{Cong Guo}, \bibinfo{person}{Bo~Yang Hsueh},
  \bibinfo{person}{Jingwen Leng}, \bibinfo{person}{Yuxian Qiu},
  \bibinfo{person}{Yue Guan}, \bibinfo{person}{Zehuan Wang},
  \bibinfo{person}{Xiaoying Jia}, \bibinfo{person}{Xipeng Li},
  \bibinfo{person}{Minyi Guo}, {and} \bibinfo{person}{Yuhao Zhu}.}
  \bibinfo{year}{2020}\natexlab{}.
\newblock \showarticletitle{Accelerating sparse dnn models without
  hardware-support via tile-wise sparsity}. In \bibinfo{booktitle}{\emph{SC20:
  International Conference for High Performance Computing, Networking, Storage
  and Analysis}}. IEEE, \bibinfo{pages}{1--15}.
\newblock


\bibitem[Guthaus et~al\mbox{.}(2016)]%
        {openram}
\bibfield{author}{\bibinfo{person}{Matthew~R. Guthaus},
  \bibinfo{person}{James~E. Stine}, \bibinfo{person}{Samira Ataei},
  \bibinfo{person}{Brian Chen}, \bibinfo{person}{Bin Wu}, {and}
  \bibinfo{person}{Mehedi Sarwar}.} \bibinfo{year}{2016}\natexlab{}.
\newblock \showarticletitle{OpenRAM: An open-source memory compiler}. In
  \bibinfo{booktitle}{\emph{2016 IEEE/ACM International Conference on
  Computer-Aided Design (ICCAD)}}. \bibinfo{pages}{1--6}.
\newblock
\urldef\tempurl%
\url{https://doi.org/10.1145/2966986.2980098}
\showDOI{\tempurl}


\bibitem[Hasinoff et~al\mbox{.}(2016)]%
        {hasinoff2016burst}
\bibfield{author}{\bibinfo{person}{Samuel~W Hasinoff}, \bibinfo{person}{Dillon
  Sharlet}, \bibinfo{person}{Ryan Geiss}, \bibinfo{person}{Andrew Adams},
  \bibinfo{person}{Jonathan~T Barron}, \bibinfo{person}{Florian Kainz},
  \bibinfo{person}{Jiawen Chen}, {and} \bibinfo{person}{Marc Levoy}.}
  \bibinfo{year}{2016}\natexlab{}.
\newblock \showarticletitle{Burst photography for high dynamic range and
  low-light imaging on mobile cameras}.
\newblock \bibinfo{journal}{\emph{ACM Transactions on Graphics (ToG)}}
  \bibinfo{volume}{35}, \bibinfo{number}{6} (\bibinfo{year}{2016}),
  \bibinfo{pages}{1--12}.
\newblock


\bibitem[Hegarty et~al\mbox{.}(2014)]%
        {hegarty2014darkroom}
\bibfield{author}{\bibinfo{person}{James Hegarty}, \bibinfo{person}{John
  Brunhaver}, \bibinfo{person}{Zachary DeVito}, \bibinfo{person}{Jonathan
  Ragan-Kelley}, \bibinfo{person}{Noy Cohen}, \bibinfo{person}{Steven Bell},
  \bibinfo{person}{Artem Vasilyev}, \bibinfo{person}{Mark Horowitz}, {and}
  \bibinfo{person}{Pat Hanrahan}.} \bibinfo{year}{2014}\natexlab{}.
\newblock \showarticletitle{Darkroom: compiling high-level image processing
  code into hardware pipelines.}
\newblock \bibinfo{journal}{\emph{ACM Trans. Graph.}} \bibinfo{volume}{33},
  \bibinfo{number}{4} (\bibinfo{year}{2014}), \bibinfo{pages}{144--1}.
\newblock


\bibitem[Hegarty et~al\mbox{.}(2016)]%
        {rigel}
\bibfield{author}{\bibinfo{person}{James Hegarty}, \bibinfo{person}{Ross Daly},
  \bibinfo{person}{Zachary DeVito}, \bibinfo{person}{Jonathan Ragan-Kelley},
  \bibinfo{person}{Mark Horowitz}, {and} \bibinfo{person}{Pat Hanrahan}.}
  \bibinfo{year}{2016}\natexlab{}.
\newblock \showarticletitle{{Rigel: Flexible Multi-Rate Image Processing
  Hardware}}. In \bibinfo{booktitle}{\emph{Proc. of SIGGRAPH}}.
\newblock


\bibitem[Holewinski et~al\mbox{.}(2012)]%
        {holewinski2012high}
\bibfield{author}{\bibinfo{person}{Justin Holewinski},
  \bibinfo{person}{Louis-No{\"e}l Pouchet}, {and} \bibinfo{person}{Ponnuswamy
  Sadayappan}.} \bibinfo{year}{2012}\natexlab{}.
\newblock \showarticletitle{High-performance code generation for stencil
  computations on GPU architectures}. In \bibinfo{booktitle}{\emph{Proceedings
  of the 26th ACM international conference on Supercomputing}}.
  \bibinfo{pages}{311--320}.
\newblock


\bibitem[Huang et~al\mbox{.}(2021)]%
        {huang2021cosa}
\bibfield{author}{\bibinfo{person}{Qijing Huang}, \bibinfo{person}{Aravind
  Kalaiah}, \bibinfo{person}{Minwoo Kang}, \bibinfo{person}{James Demmel},
  \bibinfo{person}{Grace Dinh}, \bibinfo{person}{John Wawrzynek},
  \bibinfo{person}{Thomas Norell}, {and} \bibinfo{person}{Yakun~Sophia Shao}.}
  \bibinfo{year}{2021}\natexlab{}.
\newblock \showarticletitle{Cosa: Scheduling by constrained optimization for
  spatial accelerators}. In \bibinfo{booktitle}{\emph{2021 ACM/IEEE 48th Annual
  International Symposium on Computer Architecture (ISCA)}}. IEEE,
  \bibinfo{pages}{554--566}.
\newblock


\bibitem[Koeplinger et~al\mbox{.}(2018)]%
        {koeplinger2018spatial}
\bibfield{author}{\bibinfo{person}{David Koeplinger}, \bibinfo{person}{Matthew
  Feldman}, \bibinfo{person}{Raghu Prabhakar}, \bibinfo{person}{Yaqi Zhang},
  \bibinfo{person}{Stefan Hadjis}, \bibinfo{person}{Ruben Fiszel},
  \bibinfo{person}{Tian Zhao}, \bibinfo{person}{Luigi Nardi},
  \bibinfo{person}{Ardavan Pedram}, \bibinfo{person}{Christos Kozyrakis},
  {et~al\mbox{.}}} \bibinfo{year}{2018}\natexlab{}.
\newblock \showarticletitle{Spatial: A language and compiler for application
  accelerators}. In \bibinfo{booktitle}{\emph{Proceedings of the 39th ACM
  SIGPLAN Conference on Programming Language Design and Implementation}}.
  \bibinfo{pages}{296--311}.
\newblock


\bibitem[Lai et~al\mbox{.}(2019)]%
        {lai2019heterocl}
\bibfield{author}{\bibinfo{person}{Yi-Hsiang Lai}, \bibinfo{person}{Yuze Chi},
  \bibinfo{person}{Yuwei Hu}, \bibinfo{person}{Jie Wang},
  \bibinfo{person}{Cody~Hao Yu}, \bibinfo{person}{Yuan Zhou},
  \bibinfo{person}{Jason Cong}, {and} \bibinfo{person}{Zhiru Zhang}.}
  \bibinfo{year}{2019}\natexlab{}.
\newblock \showarticletitle{HeteroCL: A multi-paradigm programming
  infrastructure for software-defined reconfigurable computing}. In
  \bibinfo{booktitle}{\emph{Proceedings of the 2019 ACM/SIGDA International
  Symposium on Field-Programmable Gate Arrays}}. \bibinfo{pages}{242--251}.
\newblock


\bibitem[Li et~al\mbox{.}(2020)]%
        {li2020heterohalide}
\bibfield{author}{\bibinfo{person}{Jiajie Li}, \bibinfo{person}{Yuze Chi},
  {and} \bibinfo{person}{Jason Cong}.} \bibinfo{year}{2020}\natexlab{}.
\newblock \showarticletitle{HeteroHalide: From image processing DSL to
  efficient FPGA acceleration}. In \bibinfo{booktitle}{\emph{Proceedings of the
  2020 ACM/SIGDA International Symposium on Field-Programmable Gate Arrays}}.
  \bibinfo{pages}{51--57}.
\newblock


\bibitem[Liu et~al\mbox{.}(2022)]%
        {liu2022s2ta}
\bibfield{author}{\bibinfo{person}{Zhi-Gang Liu}, \bibinfo{person}{Paul~N
  Whatmough}, \bibinfo{person}{Yuhao Zhu}, {and} \bibinfo{person}{Matthew
  Mattina}.} \bibinfo{year}{2022}\natexlab{}.
\newblock \showarticletitle{S2ta: Exploiting structured sparsity for
  energy-efficient mobile cnn acceleration}. In \bibinfo{booktitle}{\emph{2022
  IEEE International Symposium on High-Performance Computer Architecture
  (HPCA)}}. IEEE, \bibinfo{pages}{573--586}.
\newblock


\bibitem[Mahmoud et~al\mbox{.}(2017)]%
        {mahmoud2017ideal}
\bibfield{author}{\bibinfo{person}{Mostafa Mahmoud}, \bibinfo{person}{Bojian
  Zheng}, \bibinfo{person}{Alberto~Delm{\'a}s Lascorz}, \bibinfo{person}{Felix
  Heide}, \bibinfo{person}{Jonathan Assouline}, \bibinfo{person}{Paul Boucher},
  \bibinfo{person}{Emmanuel Onzon}, {and} \bibinfo{person}{Andreas Moshovos}.}
  \bibinfo{year}{2017}\natexlab{}.
\newblock \showarticletitle{IDEAL: Image denoising accelerator}. In
  \bibinfo{booktitle}{\emph{2017 50th Annual IEEE/ACM International Symposium
  on Microarchitecture (MICRO)}}. IEEE, \bibinfo{pages}{82--95}.
\newblock


\bibitem[Murray et~al\mbox{.}(2016)]%
        {murray2016microarchitecture}
\bibfield{author}{\bibinfo{person}{Sean Murray}, \bibinfo{person}{William
  Floyd-Jones}, \bibinfo{person}{Ying Qi}, \bibinfo{person}{George Konidaris},
  {and} \bibinfo{person}{Daniel~J Sorin}.} \bibinfo{year}{2016}\natexlab{}.
\newblock \showarticletitle{{The microarchitecture of a real-time robot motion
  planning accelerator}}. In \bibinfo{booktitle}{\emph{The 49th Annual IEEE/ACM
  International Symposium on Microarchitecture}}. IEEE Press,
  \bibinfo{pages}{45}.
\newblock


\bibitem[Nigam et~al\mbox{.}(2020)]%
        {nigam2020predictable}
\bibfield{author}{\bibinfo{person}{Rachit Nigam}, \bibinfo{person}{Sachille
  Atapattu}, \bibinfo{person}{Samuel Thomas}, \bibinfo{person}{Zhijing Li},
  \bibinfo{person}{Theodore Bauer}, \bibinfo{person}{Yuwei Ye},
  \bibinfo{person}{Apurva Koti}, \bibinfo{person}{Adrian Sampson}, {and}
  \bibinfo{person}{Zhiru Zhang}.} \bibinfo{year}{2020}\natexlab{}.
\newblock \showarticletitle{Predictable accelerator design with time-sensitive
  affine types}. In \bibinfo{booktitle}{\emph{Proceedings of the 41st ACM
  SIGPLAN Conference on Programming Language Design and Implementation}}.
  \bibinfo{pages}{393--407}.
\newblock


\bibitem[Nowatzki et~al\mbox{.}(2013)]%
        {nowatzki2013general}
\bibfield{author}{\bibinfo{person}{Tony Nowatzki}, \bibinfo{person}{Michael
  Sartin-Tarm}, \bibinfo{person}{Lorenzo De~Carli},
  \bibinfo{person}{Karthikeyan Sankaralingam}, \bibinfo{person}{Cristian
  Estan}, {and} \bibinfo{person}{Behnam Robatmili}.}
  \bibinfo{year}{2013}\natexlab{}.
\newblock \showarticletitle{{A general constraint-centric scheduling framework
  for spatial architectures}}.
\newblock \bibinfo{journal}{\emph{ACM SIGPLAN Notices}} \bibinfo{volume}{48},
  \bibinfo{number}{6} (\bibinfo{year}{2013}), \bibinfo{pages}{495--506}.
\newblock


\bibitem[Pu et~al\mbox{.}(2017)]%
        {pu2017programming}
\bibfield{author}{\bibinfo{person}{Jing Pu}, \bibinfo{person}{Steven Bell},
  \bibinfo{person}{Xuan Yang}, \bibinfo{person}{Jeff Setter},
  \bibinfo{person}{Stephen Richardson}, \bibinfo{person}{Jonathan
  Ragan-Kelley}, {and} \bibinfo{person}{Mark Horowitz}.}
  \bibinfo{year}{2017}\natexlab{}.
\newblock \showarticletitle{Programming heterogeneous systems from an image
  processing DSL}.
\newblock \bibinfo{journal}{\emph{ACM Transactions on Architecture and Code
  Optimization (TACO)}} \bibinfo{volume}{14}, \bibinfo{number}{3}
  (\bibinfo{year}{2017}), \bibinfo{pages}{1--25}.
\newblock


\bibitem[Ragan-Kelley et~al\mbox{.}(2013)]%
        {ragan2013halide}
\bibfield{author}{\bibinfo{person}{Jonathan Ragan-Kelley},
  \bibinfo{person}{Connelly Barnes}, \bibinfo{person}{Andrew Adams},
  \bibinfo{person}{Sylvain Paris}, \bibinfo{person}{Fr{\'e}do Durand}, {and}
  \bibinfo{person}{Saman Amarasinghe}.} \bibinfo{year}{2013}\natexlab{}.
\newblock \showarticletitle{Halide: a language and compiler for optimizing
  parallelism, locality, and recomputation in image processing pipelines}.
\newblock \bibinfo{journal}{\emph{Acm Sigplan Notices}} \bibinfo{volume}{48},
  \bibinfo{number}{6} (\bibinfo{year}{2013}), \bibinfo{pages}{519--530}.
\newblock


\bibitem[Ranganathan et~al\mbox{.}(2021)]%
        {ranganathan2021warehouse}
\bibfield{author}{\bibinfo{person}{Parthasarathy Ranganathan},
  \bibinfo{person}{Daniel Stodolsky}, \bibinfo{person}{Jeff Calow},
  \bibinfo{person}{Jeremy Dorfman}, \bibinfo{person}{Marisabel Guevara},
  \bibinfo{person}{Clinton~Wills Smullen~IV}, \bibinfo{person}{Aki Kuusela},
  \bibinfo{person}{Raghu Balasubramanian}, \bibinfo{person}{Sandeep Bhatia},
  \bibinfo{person}{Prakash Chauhan}, {et~al\mbox{.}}}
  \bibinfo{year}{2021}\natexlab{}.
\newblock \showarticletitle{Warehouse-scale video acceleration: co-design and
  deployment in the wild}. In \bibinfo{booktitle}{\emph{Proceedings of the 26th
  ACM International Conference on Architectural Support for Programming
  Languages and Operating Systems}}. \bibinfo{pages}{600--615}.
\newblock


\bibitem[Redgrave et~al\mbox{.}(2018)]%
        {redgrave2018pixel}
\bibfield{author}{\bibinfo{person}{Jason Redgrave}, \bibinfo{person}{Albert
  Meixner}, \bibinfo{person}{Nathan Goulding-Hotta}, \bibinfo{person}{Artem
  Vasilyev}, {and} \bibinfo{person}{Ofer Shacham}.}
  \bibinfo{year}{2018}\natexlab{}.
\newblock \showarticletitle{Pixel Visual Core: Google’s Fully
  ProgrammableImage, Vision, and AI Processor For Mobile Devices}. In
  \bibinfo{booktitle}{\emph{Proc. IEEE Hot Chips Symp.(HCS)}}.
  \bibinfo{pages}{1--18}.
\newblock


\bibitem[Sacks et~al\mbox{.}(2018)]%
        {sacks2018robox}
\bibfield{author}{\bibinfo{person}{Jacob Sacks}, \bibinfo{person}{Divya
  Mahajan}, \bibinfo{person}{Richard~C Lawson}, {and} \bibinfo{person}{Hadi
  Esmaeilzadeh}.} \bibinfo{year}{2018}\natexlab{}.
\newblock \showarticletitle{{Robox: an end-to-end solution to accelerate
  autonomous control in robotics}}. In \bibinfo{booktitle}{\emph{Proceedings of
  the 45th Annual International Symposium on Computer Architecture}}. IEEE
  Press, \bibinfo{pages}{479--490}.
\newblock


\bibitem[Stewart et~al\mbox{.}(2018)]%
        {stewart2018ripl}
\bibfield{author}{\bibinfo{person}{Robert Stewart}, \bibinfo{person}{Kirsty
  Duncan}, \bibinfo{person}{Greg Michaelson}, \bibinfo{person}{Paulo Garcia},
  \bibinfo{person}{Deepayan Bhowmik}, {and} \bibinfo{person}{Andrew Wallace}.}
  \bibinfo{year}{2018}\natexlab{}.
\newblock \showarticletitle{RIPL: A parallel image processing language for
  FPGAs}.
\newblock \bibinfo{journal}{\emph{ACM Transactions on Reconfigurable Technology
  and Systems (TRETS)}} \bibinfo{volume}{11}, \bibinfo{number}{1}
  (\bibinfo{year}{2018}), \bibinfo{pages}{1--24}.
\newblock


\bibitem[Suleiman et~al\mbox{.}(2019)]%
        {suleiman2019navion}
\bibfield{author}{\bibinfo{person}{Amr Suleiman}, \bibinfo{person}{Zhengdong
  Zhang}, \bibinfo{person}{Luca Carlone}, \bibinfo{person}{Sertac Karaman},
  {and} \bibinfo{person}{Vivienne Sze}.} \bibinfo{year}{2019}\natexlab{}.
\newblock \showarticletitle{Navion: A 2-mW Fully Integrated Real-Time
  Visual-Inertial Odometry Accelerator for Autonomous Navigation of Nano
  Drones}.
\newblock \bibinfo{journal}{\emph{IEEE Journal of Solid-State Circuits}}
  \bibinfo{volume}{54}, \bibinfo{number}{4} (\bibinfo{year}{2019}),
  \bibinfo{pages}{1106--1119}.
\newblock


\bibitem[Vasilyev et~al\mbox{.}(2016)]%
        {paisp}
\bibfield{author}{\bibinfo{person}{Artem Vasilyev}, \bibinfo{person}{Nikhil
  Bhagdikar}, \bibinfo{person}{Ardavan Pedram}, \bibinfo{person}{Stephen
  Richardson}, \bibinfo{person}{Shahar Kvatinsky}, {and} \bibinfo{person}{Mark
  Horowitz}.} \bibinfo{year}{2016}\natexlab{}.
\newblock \showarticletitle{{Evaluating Programmable Architectures for Imaging
  and Vision Applications}}. In \bibinfo{booktitle}{\emph{Proc. of MICRO}}.
\newblock


\bibitem[Weng et~al\mbox{.}(2020)]%
        {weng2020dsagen}
\bibfield{author}{\bibinfo{person}{Jian Weng}, \bibinfo{person}{Sihao Liu},
  \bibinfo{person}{Vidushi Dadu}, \bibinfo{person}{Zhengrong Wang},
  \bibinfo{person}{Preyas Shah}, {and} \bibinfo{person}{Tony Nowatzki}.}
  \bibinfo{year}{2020}\natexlab{}.
\newblock \showarticletitle{Dsagen: Synthesizing programmable spatial
  accelerators}. In \bibinfo{booktitle}{\emph{2020 ACM/IEEE 47th Annual
  International Symposium on Computer Architecture (ISCA)}}. IEEE,
  \bibinfo{pages}{268--281}.
\newblock


\bibitem[Weste and Harris(2015)]%
        {weste2015cmos}
\bibfield{author}{\bibinfo{person}{Neil~HE Weste} {and} \bibinfo{person}{David
  Harris}.} \bibinfo{year}{2015}\natexlab{}.
\newblock \bibinfo{booktitle}{\emph{CMOS VLSI design: a circuits and systems
  perspective}}.
\newblock \bibinfo{publisher}{Pearson Education India}.
\newblock


\bibitem[Whatmough et~al\mbox{.}(2019)]%
        {whatmough2019fixynn}
\bibfield{author}{\bibinfo{person}{Paul~N Whatmough}, \bibinfo{person}{Chuteng
  Zhou}, \bibinfo{person}{Patrick Hansen}, \bibinfo{person}{Shreyas~Kolala
  Venkataramanaiah}, \bibinfo{person}{Jae-sun Seo}, {and}
  \bibinfo{person}{Matthew Mattina}.} \bibinfo{year}{2019}\natexlab{}.
\newblock \showarticletitle{Fixynn: Efficient hardware for mobile computer
  vision via transfer learning}.
\newblock \bibinfo{journal}{\emph{arXiv preprint arXiv:1902.11128}}
  (\bibinfo{year}{2019}).
\newblock


\bibitem[Xu et~al\mbox{.}(2019)]%
        {xu2019tigris}
\bibfield{author}{\bibinfo{person}{Tiancheng Xu}, \bibinfo{person}{Boyuan
  Tian}, {and} \bibinfo{person}{Yuhao Zhu}.} \bibinfo{year}{2019}\natexlab{}.
\newblock \showarticletitle{Tigris: Architecture and algorithms for 3d
  perception in point clouds}. In \bibinfo{booktitle}{\emph{Proceedings of the
  52nd Annual IEEE/ACM International Symposium on Microarchitecture}}.
  \bibinfo{pages}{629--642}.
\newblock


\bibitem[Zhu et~al\mbox{.}(2018)]%
        {zhu2018euphrates}
\bibfield{author}{\bibinfo{person}{Yuhao Zhu}, \bibinfo{person}{Anand
  Samajdar}, \bibinfo{person}{Matthew Mattina}, {and} \bibinfo{person}{Paul
  Whatmough}.} \bibinfo{year}{2018}\natexlab{}.
\newblock \showarticletitle{Euphrates: Algorithm-SoC Co-Design for Low-Power
  Mobile Continuous Vision}. In \bibinfo{booktitle}{\emph{Proceedings of the
  45th ACM/IEEE Annual International Symposium on Computer Architecture}}.
\newblock


\end{thebibliography}

\end{document}